\pdfoutput=1
\documentclass[11pt]{article}
\usepackage[affil-it]{authblk} % Package for author affiliation
\usepackage[authoryear]{natbib}
\usepackage{gensymb}
\usepackage{amsmath}
\usepackage{amssymb}
\usepackage{amsfonts}
\usepackage{textcomp}
\usepackage{graphicx}
\usepackage{float}
\usepackage{booktabs}
\usepackage[a4paper, top=1in,bottom=1in,left=1in,right=1in,marginparwidth=0cm]{geometry}
\usepackage[title]{appendix}
\usepackage{pdflscape}
\usepackage{longtable}

\usepackage{adjustbox}
\usepackage{array}
\usepackage{caption}
\usepackage{chngcntr}
\usepackage{mathtools}
\usepackage{lineno}
\usepackage{setspace}
\usepackage{makecell}
\usepackage{hyperref}
\hypersetup{colorlinks=true,allcolors=blue}
\usepackage{algorithm}
\usepackage{algorithmic}
\usepackage{accents}
\usepackage{mathrsfs}

\newcolumntype{R}[2]{%
    >{\adjustbox{angle=#1,lap=\width-(#2)}\bgroup}%
    l%
    <{\egroup}%
}
\newcommand*\rot{\multicolumn{1}{R{45}{1em}}}% no optional argument here, please!

\newcommand{\citepos}[1]{\citeauthor{#1}'s (\citeyear{#1})}
\renewcommand{\eqref}[1]{Eq.\,\ref{#1}}

\newcommand{\myvect}[1]{\accentset{\rightharpoonup}{#1}}

\title{Negligible effects of environmental fluctuations on the maintenance of coral biodiversity: A test of five storage effects}
\author[1]{Evan C. Johnson}
\author[1,2]{Sean R. Connolly}
\affil[1]{Naos Marine Laboratories, Smithsonian Tropical Research Institute, Ancón, Panama}
\affil[2]{College of Science and Engineering, James Cook University, Townsville, Queensland, Australia}
\affil[*]{Corresponding author: Evan Johnson, JohnsonE@si.edu}

\date{}

\begin{document}
\maketitle 

\newpage

\tableofcontents

\newpage 

\section*{Abstract}

The storage effect is a general explanation for ecological coexistence, wherein different species specialize on different states of a fluctuating environment, e.g., hot vs. cold years. Despite the storage effect’s prominence in theoretical ecology, we lack evidence on whether it maintains biodiversity in nature. Here, we examine five storage effects in a community of 11 coral species from the Great Barrier Reef, using detailed size-structured demographic data collected over five years. We parameterize integral projection models, simulate coral communities, and quantify coexistence mechanisms through Modern Coexistence Theory. Results show that storage effects promote coexistence but are weak compared to fitness differences. Despite coral communities exhibiting theoretical prerequisites for strong temporal niche partitioning, the storage effect plays only a minor role in maintaining coral biodiversity. This aligns with growing evidence that storage effects are weak across ecosystems. Coral coexistence likely depends more on spatial processes, including microhabitat partitioning and asymmetric dispersal.

\newpage 

\section{Introduction} \label{Introduction}

Explaining coexistence in species-rich ecosystems is a fundamental problem in ecology. According to classic ecological theory, an environment can only support as many species as there are discrete limiting factors, such as resources or natural enemies \citep{volterra1926variationsST, Gause1934, levin1970community}. However, in many high-diversity systems --- including coral reef fishes, tropical trees, Mediterranean grasslands, and Australian shrublands --- hundreds of species co-occur within a square kilometer \citep{condit1996species, cowling1996plant, veron2011coral}, despite having only a handful of recognizable limiting factors upon which to specialize.

The storage effect \citep{chesson1981environmentalST, chesson1994multispecies} offers a resolution to this phenomenon, termed 'the paradox of the plankton', by enabling species coexistence through temporal niche partitioning, rather than specialization on limited discrete resources. To illustrate, consider two coral-dwelling goby species, \textit{Gobiodon histrio} and \textit{Gobiodon erythrospilus}, that respectively produce many larvae during warm and cool years. The larvae of these species appear to compete in a lottery for shelter in unoccupied coral heads of particular host species \citep{munday2004competitive}, and relationships between abundances and latitudinal ranges suggests overlapping but distinct thermal niches \citep{munday1999gobiodon}. When an abundant species experiences a good year, it produces many larvae that compete fiercely for available coral host colonies. Thus, favorable environmental conditions are undermined by the high level of competition that they bring about. This disadvantage doesn't affect rare species because, while their per capita larval production may be high in their good years, the absolute number of larvae remains low due to the rarity of adult gobies. This creates a rare-species advantage --- the hallmark of stable coexistence. 

To understand how the storage effect applies beyond goby species and fecundity fluctuations, we can refer to its ingredient-list definition \citep{chesson1994multispecies, johnson2022towards}. This formulation states that the storage effect generally promotes coexistence among multiple species when three key elements are present: 1) species-specific responses to a variable environment, 2) covariance between environment and competition, and 3) non-additivity, which is essentially an interaction effect between environment and competition on per capita growth rates, such that the effects of competition are more intense when the environment is favorable. Though these ingredients are abstract, they directly map onto our goby example. The two goby species prefer different temperatures (species-specific environmental responses), a favorable environment for the common species generates high competition (environment-competition covariance), and the simultaneous occurrence of high larval production and intense competition for space negatively impacts growth (non-additivity).

Despite the storage effect's potential for explaining coexistence, we do not know how important it is in real communities. Some arguments suggest the ubiquity of the storage effect. Non-additivity (ingredient \# 3) arises in all but the simplest models \citep{chesson1994multispecies, adler2006climate}, and species-specific responses to the environment (ingredient \#2) are substantiated by two lines of evidence: recruitment fluctuations often vary by 2--3 orders of magnitude \citep{robles1997changing, beukema2007variability, houde2016recruitment} and species' per capita demographic rates are typically weakly correlated \citep{pake1996seed, usinowicz2017temporal, ruiz2024high}.

Several theoretical arguments suggest the storage effect may be generally weak (as detailed by \citealp{stump2023reexamining}). Additionally, multiple theoretical analyses indicate that spatial coexistence mechanisms more reliably support coexistence because species can remain in their good patches, compounding  benefits of favorable conditions over multiple generations --- assuming dispersal and undirected movement are not too extensive relative to patch size \citep{snyder2003local, snyder2004spatial} --- whereas species cannot remain in their ``good years'' (\citealp{chesson1985coexistence, snyder2008does}; but see \citealp{ellner2024s}), instead suffering exponential decline and possible extinction during unfavorable periods \citep{adler2008environmental, pande2020mean}.

Empirical evidence for the storage effect is mixed and context-dependent. Laboratory experiments confirm that the storage effect can drive coexistence under controlled conditions \citep{jiang2007temperature, descamps2005stable}, but field studies present a more complex picture. Many investigations have documented the presence of storage effect ingredients without quantifying the mechanism's relative importance \citep{venable1993diversity, pake1995coexistence, pake1996seed, kelly2002coexistence, facelli2005differences, kelly2005new, chesson2012storage, holt2014variation}. Some models, parameterized with data from natural ecosystems, find that temporal fluctuations are essential for coexistence \citep{angert2009functional, usinowicz2012coexistence, caceres1997temporal, adler2006climate}. However, when researchers simultaneously quantify multiple coexistence mechanisms, the storage effect consistently appears weak relative to other coexistence mechanisms or fluctuation-independent fitness differences \citep{armitage2019negative, armitage2020coexistence, hallett2019rainfall, zepeda2019fluctuation, letten2019mechanistic, Ellner2019}. Recently, \citet{stump2023reexamining} analyzed communities previously thought to exhibit strong storage effects (annual plants, tropical trees, and iguanid lizards) and found that the storage effect either undermined coexistence or contributed positively but weakly, with effects 34-73\% smaller than those of resource or predator partitioning.

Previous literature on the storage effect presents four noteworthy limitations. First, measurements of the storage effect typically rely on models that may not capture ecological complexity, and rarely include sensitivity analyses to test robustness to simplifying assumptions.  Second, most studies fail to propagate parameter uncertainty to coexistence outcomes, obscuring whether findings reflect ecological reality or modeling/sampling-error artifacts. Third, research has disproportionately focused on fecundity or recruitment fluctuations, while neglecting potential storage effects arising from variation in survival or growth rates. (but see \citep{alvarez2023disturbance}; \citealp{adler2006climate}). Finally, limited temporal replication in field studies may prevent the detection of of rare 'jackpot years' when environmental fluctuations dramatically alter demographic rates by orders of magnitude (e.g., \citealp{harper1977population}).

With characteristics thought to favor the storage effect, coral communities are ideal for providing  ``existence proof'' evidence that the the storage effect is important in some ecosystems. Alternatively,  minimal storage effects in corals would suggest the limited importance of this coexistence mechanism across ecological systems. Coral communities exhibit large recruitment variability at all spatiotemporal scales \citep{dunstan1998spatio, glassom2004coral, nakamura2010spatiotemporal, edmunds2021spatiotemporal}, providing evidence for ingredient \#1 (species-specific environmental responses); though proving species-specificity remains challenging because coral recruits are difficult to identify at the species level \citep{babcock2003identification}. Adult coral longevity contributes to ingredient \#3, non-additivity \citep{muko2000species, snyder2003local}.

\begin{figure}[H]
\centering
\includegraphics[scale = 1]{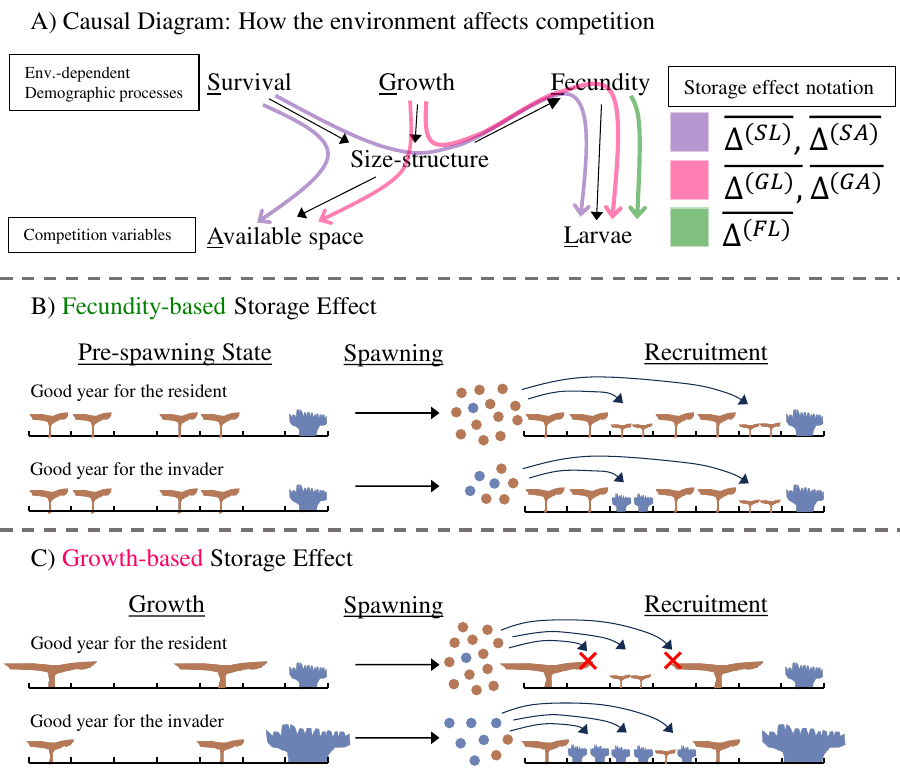}
\caption{Illustration of the 5 storage effects (SE). \textbf{Panel A)} A causal diagram demonstrates how a good environment leads to high competition for common species, a key ingredient in SEs. Crucially, several SEs are mediated through population size-structure. The symbols ($\overline{\Delta^{(SL)}}$, $\overline{\Delta^{(SA)}}$, $\overline{\Delta^{(GL)}}$, $\overline{\Delta^{(GA)}}$, $\overline{\Delta^{(FL)}}$) represent the 5 distinct community-average storage effects, which measure the effects of covariance between environmental factors (\underline{S}urvival, \underline{G}rowth, \underline{F}ecundity) and competition types (\underline{L}arvae, \underline{A}rea); see Section \ref{Modern coexistence theory} for mathematical details. \textbf{Panel B)} An example of the SE due to fecundity fluctuations. When the resident experiences a good year, many larvae are produced (visualized as a large cluster of circles in the figure), leading to high competition. When the rare invader experiences a good year, many larvae are produced \textit{per invader colony}, but total competition is lower by dint of the invader's rarity. \textbf{Panel C)} An example of the SE due to fluctuations in colony growth rates. When the resident experiences a good year of growth, competition increases in two ways: large adult colonies produce more larvae, and large adults occupy space, thus preventing recruitment (visualized as red crosses in the figure). In comparison, a good year of growth for the invader contributes less to adult colony cover, and to the total larvae supply, and thus does not produce as much competition.}
\label{fig:conceptual_fig}
\end{figure}

Here, we quantify the storage effect in a diverse coral community on the Great Barrier Reef using detailed, size-structured demographic data (11 species over 5 years), Integral Projection Models (IPMs), and Modern Coexistence Theory (MCT). The model generates five types of storage effects (Fig. \ref{fig:conceptual_fig}) based on three environmentally-dependent demographic processes (survival, growth, and fecundity) and two measures of competition (adult colony cover and larval densities). Importantly, we directly addresses the four key limitations identified in previous research. Specifically, we: 1) conduct many simulations to assess sensitivity to model assumptions; 2) employ a Bayesian framework to explicitly propagate parameter uncertainty to coexistence outcomes; 3) investigate multiple pathways through which the storage effect can operate, arising from variability in survival and growth in addition to fecundity; and 4) exploring the potential impact of unobserved 'jackpot years' by virtually increasing environmental variability, and by comparing the magnitude of environmental variation in our dataset to other sources in the literature.

By providing this robust quantitative assessment in a system potentially conducive to strong temporal niche partitioning, our findings contribute directly to the debate on the general importance of the storage effect. Is the storage effect generally weak, merely an academic exercise in how stochasticity can promote diversity, or is it the main driver of coexistence in real ecosystems? We find that while storage effects do operate in this system, they are generally weak and not sufficient for coexistence.

\section{Methods}

Using coral demographic data, we parameterized an Integral Projection Model (IPM) with four sequential processes: mortality, growth, spawning, and density-dependent recruitment. Parameters were estimated using Bayesian methods, with species-specific year effects in growth and fecundity capturing environmental variation. We conducted simulations across several community modules, including the full 11-species community and pairs of morphologically similar species. To identify mechanisms driving coexistence, we applied simulation-based Modern Coexistence Theory, which involves simulations of IPM with species perturbed to low density. Put together, these methods allow us to quantify five distinct storage effects.

\subsection{Data}

We used a 5-year dataset of coral demography \citep{madin2023six} from the reef crest of Lizard Island in the northern Great Barrier Reef, Australia (14.70007°S, 145.44877°E). The study tracked approximately 30 colonies from each of 11 coral species. The species were classified by their morphologies: massive (\textit{Goniastrea pectinata} and \textit{G. retiformis}), digitate (\textit{Acropora humilis} and \textit{A.} cf. \textit{digitifera}), corymbose (\textit{A. millepora}, \textit{A. nasuta}, and \textit{A. spathulata}), tabular (\textit{A. cytherea} and \textit{A. hyacinthus}), and arborescent (\textit{A. robusta} and \textit{A. intermedia}).

Colonies were photographed and their planar areas were estimated. Colony mortality was monitored, and dead colonies were replaced with newly tagged ones to maintain consistent sample sizes. Additional reproductive data came from sampling 30 fragments from nearby untagged colonies of each species, examining them for gravid status, and counting the number of eggs. The study concluded earlier than planned in early 2015 when Tropical Cyclone Nathan caused $> 90\%$ mortality of tagged colonies.

\subsection{Model description}

We modeled coral dynamics using a size-structured Integral Projection Model (IPM). Each time step represents a year and includes four sequential processes: mortality, growth, spawning, and density-dependent recruitment. The full model can be written as
\begin{equation}
    n_{j,t+1}(y) = \int_{-\infty}^{\infty} G_{j,t}(y|x)S_{j,t}(x)n_{j,t}(x)\,dx + R_{j,t} \cdot \phi(y),
\end{equation}
where $n_{j,t}(x)$ is colony density at size $x$ and time $t$, $G_{j,t}(y|x)$ is the growth kernel describing transition probabilities from size $x$ to $y$, $S_{j,t}(x)$ is the annual survival probability, and $R_{j,t} \cdot \phi(y)$ is recruitment into size class $y$ with $R_{j,t}$ representing the probability of recruitment and $\phi(y)$ the recruit size distribution. Mathematical details and model justifications are provided in Appendix \ref{Model description}. Figure \ref{fig:vital_rates_combined} summarizes the relationship between demographic processes and colony size.

\begin{figure}[H]
\centering
\makebox[\textwidth]{\includegraphics[scale=0.9]{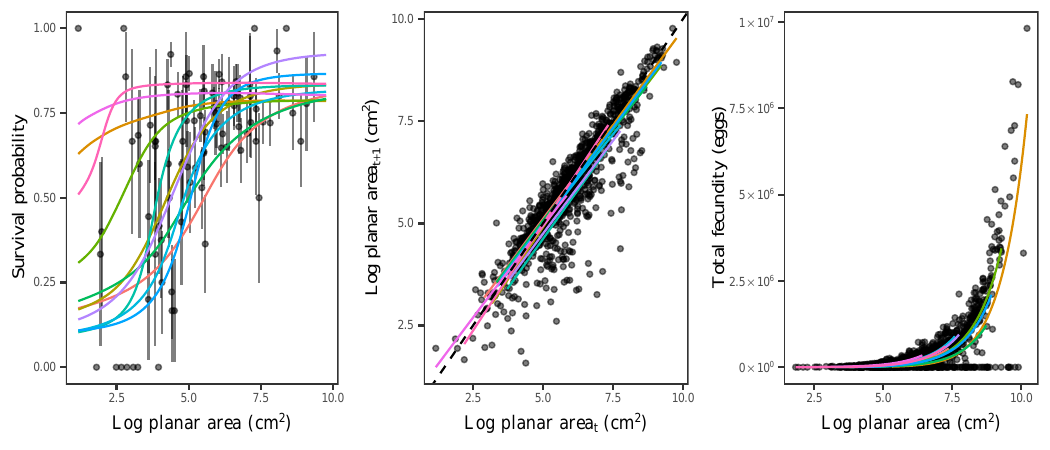}}
\caption{Visual summary of coral size-structured dynamics. Panel A) Background survival is a logistically increasing function of size. Panel B) Smaller corals tend to grow, but larger corals tend to shrink. Growth here represents net annual change, reflecting skeletal extension offset by any partial mortality losses. Panel C) Total fecundity increases rapidly with colony size. Different colors show the posterior mean relationships for different species.}
\label{fig:vital_rates_combined}
\end{figure}

An aspatial IPM reasonably approximates coral population dynamics despite local colony interactions (via overgrowth, digestion, defensive stinging cells, and competition for light \citealp{lang1990competition}). Two key observations support this approach. First, competitive interactions between colonies minimally impact growth and survival \citep{connell2004long, lapid2006long, chornesky1991ties, peach1999sweeper, alvarez2018negligible}, likely because disturbances and algae keep coral cover well below 100\%, allowing colonies to grow into open space \citep{romano1990long}, and because some overtopped colonies persist through heterotrophic feeding and diffuse light \citep{stimson1985effect}. Second, all studied species are broadcast spawners whose larvae rarely settle near parent colonies. These factors justify our aspatial approach since growth, survival, and recruitment processes don't strongly depend on spatial relationships between individuals. Our model intentionally excludes exogenous sources of spatial variation to focus specifically on evaluating the storage effect as a coexistence mechanism.

Although the storage effect requires environmental fluctuations, we don't directly measure abiotic variables since it is unclear what variables matter and how exactly they relate to coral demography. Instead, we estimate temporal variances and between-species covariances in environmentally-dependent demographic rates (survival, growth, and fecundity). Throughout the paper, "the environment" or "environmental responses" refers to these processes.

In the model, colony survival depends on two mortality sources: wave-disturbance and background mortality (Appendices \ref{Wave action survival}, \ref{Background mortality}). Wave-disturbance mortality occurs when a colony's hydrodynamic vulnerability exceeds the disturbance intensity, affecting larger and more top-heavy colonies disproportionately. This sub-model draws from previous work using historical wind and wave data from Lizard Island \citep{madin2006scaling, madin2012calcification, alvarez2023disturbance}.

Background mortality encompasses all non-wave mortality sources and decreases logistically with colony size. This function remains constant across years since environmental variability is already captured in the wave-disturbance component. Overall survival probability combines both mortality sources: $S_{j,t}(x) = \left(1- M_j(x)\right)\left(1- D_{j,t}(x)\right)$, where $M_j(x)$ is the background mortality rate and $D_{j,t}(x)$ is the wave-disturbance mortality rate for species $j$ at time $t$ for a colony of size $x$.

Colony growth depends on colony area and species-specific year effects (Appendix \ref{Colony growth}). Colony fecundity is calculated by multiplying three factors: the number of eggs per polyp, polyp density, and colony planar area (Appendix \ref{Fecundity}). Polyp densities are estimated species-specific constants, but the number of eggs per polyp depends on colony size and species-specific year effects.

Recruitment is density-dependent, depending both on the availability of free space, and the total number of larval competitors (Appendix \ref{Recruitment}). Let $L_{j,t}=\int n_{j,t}(x)F_{j,t}(x)\,dx$ be the number of eggs produced by species $j$, and let $N'_{j,t} = \int_{-\infty}^{\infty} n'_{j,t}(x) \exp(x)\,dx$ be the total adult colony cover belonging to species $j$ \textit{after} the growth and survival processes; Note that $x$ is the logarithm of planar area, so $\exp(x)$ the raw planar area. With these quantities defined, we can express the density of one-year recruits as
\begin{equation} \label{eq:recruit0}
R_{j,t+1} = \overbrace{\frac{\beta_{j,R} L_{j,t}}{\sum_{k=1}^{S} L_{k,t}}}^{\mathclap{\text{\shortstack{Recruits per m$^2$ \\ on available substrate}}}} \times \underbrace{\left(1- \sum_{j=1}^{11} N'_{j,t}\right)}_{\mathclap{\text{Proportion substrate available}}},
\end{equation}
where $S$ is the number of species, and $\beta_{j,R}$ determines the maximum recruit density. Lacking the data to estimate $\beta_{j,R}$ directly, we used simulations to identify values generating realistic coral cover (0.1-0.5) for each posterior parameter draw. These values represent typical regional cover \citep{anderson2025recovery} and align with recruitment data from a recent meta-analysis (\citealp{edmunds2023coral}; figures \ref{fig:max_recruit_dist}--\ref{fig:recruit_ecdf}). Alternative approaches for selecting $\beta_{j,R}$ values produced qualitatively identical results (Appendix \ref{Robustness analysis}).

The density-dependent nature of coral recruitment is supported by previous research. Observational data shows a negative relationship between adult coral cover and recruitment \citep{edmunds2004effects, edmunds2018implications, couch2023ecological}; and experiments show that settler survival is a decreasing function of settler density \citep{suzuki2012optimal, edwards2015direct, doropoulos2017density, cameron2020density}. The recruitment dynamics in \eqref{eq:recruit1} implicitly contain two assumptions. First, settlement probabilities are approximately equal across species, such that they cancel out in the quotient term. Second, settler abundance (and thus competition between settlers) is sufficiently high that total recruitment in open space (summed across species) is at its asymptotic maximum ---technically, a weighted average of species-specific maximum recruit densities $\beta_{j,R}$, where the weights depend on the relative larval supply from each species. These assumptions offers several advantages: they allow us to avoid specifying uncertain settlement probabilities (Appendix \ref{Early life history transition probabilities}) and enable defining competition with a single quantity (see in \eqref{eq:comp1} in the following subsection). Additionally, by assuming density-dependent recruitment in all years, we maximize the potential for storage effects in our system. This ultimately strengthens our conclusions: if we detect only weak storage effects under theoretically favorable conditions, we can be even more confident that storage effects is small. Nevertheless, we relax the assumption of consistent density-dependent recruitment in Appendix \ref{Robustness analysis} and find qualitatively identical results.
 
\subsection{Model-fitting}

We estimated parameters for growth, survival, and fecundity sub-models using Bayesian hierarchical methods in Stan \citep{carpenter2017stan}, with weakly informative priors. A standard suite of diagnostics was monitored to ensure model convergence and sampling efficiency, and the posterior contraction statistic \citep{schad2021toward} was computed to check for outsized prior influence. 

The IPM was implemented numerically by discretizing log planar area into bins, with integrals evaluated using the midpoint rule. To address eviction --- probability mass leaving the model due to limits on size classes \citep{williams2012avoiding} --- we implemented a correction procedure that reassigned probability mass that would fall outside the boundaries to the respective minimum or maximum bins. More details are provided in \ref{Model-fitting details}--\ref{IPM implementation}.

Correlations between species' year effects were modeled by estimating species-specific year effects and then computing the empirical correlation matrix from these estimates. This approach captures biologically realistic patterns where different species pairs exhibit varying degrees of correlation, but it may overestimate variability due to limited temporal replication. In Appendix \ref{Robustness analysis}, we explored an alternative ``diffuse correlation'' approach where all species pairs share the same correlation coefficient for each demographic process. Both methods yielded qualitatively identical results.

\subsection{Simulations}

We conducted simulations across several distinct \textit{community modules}. The first module contained all 11 coral species, allowing us to assess the potential coexistence of many species. The second module focused specifically on the codominant species (\textit{A. digitifera} and \textit{A. hyacinthus}) which together constitute about 45\% of individuals (Data in \citealp{dornelas2008multiple}). Additionally, we simulated paired species within each of the five colony growth form types (arborescent, tabular, corymbose, digitate, and massive) to examine coexistence mechanisms between morphologically similar species and to detect coexistence mechanisms that might be obscured if one species dominates in the full-community module. 

For each community module, we sampled 200 parameter sets from the joint posterior. Initial community assembly began with all species in the module at equal cover (total initial cover of 50\%), followed by 10 years of simulation with minimal immigration (2 recruits m\textsuperscript{-2} per species per year). We then continued simulations for 1000 additional years without immigration to determine which species could persist in the long term. Species were considered extirpated if their cover dropped below 10\textsuperscript{-10} at any point during the simulation, which is approximately 2 orders of magnitude lower than the smallest relative abundances in historical surveys \citep{dornelas2008multiple, mcwilliam2023net}.

\subsection{Modern Coexistence Theory} \label{MCT_methods}

To determine the mechanisms driving coexistence in our coral reef system, we used simulation-based Modern Coexistence Theory (MCT; \citealp{ellner2016data, Ellner2019, johnson2023coexistence}). This approach decomposes species' per capita growth rates when rare (invasion growth rates) into interpretable additive components, and then compares components between a species perturbed to low density (the invader), and species at their typical abundances (residents). Because a rare-species advantage is the hallmark of stable coexistence, these invader--resident comparisons are called \textit{coexistence mechanisms}, with the storage effect being one such mechanism. 

For each species, we calculated the per capita growth rate, defined as the natural logarithm of the quotient of coral cover in successive years. We then decomposed the temporal average of this rate as
\begin{equation} \label{eq:coarseMCT}
\overline{r_j(E_j,C_j)} = \varepsilon_j^0 + \varepsilon_j^E + \varepsilon_j^C + \varepsilon_j^{EC}
\end{equation}
In this equation, $E_j$ represents the set of environmentally dependent parameters (i.e., the dislodgement mechanical threshold, as well as year effects in growth and fecundity). The competition parameter $C$ is defined as the logarithm of total egg production relative to available space:
\begin{equation} \label{eq:comp1}
C_{t} = \log\left(\frac{\sum_{k=1}^{11} L_{k,t}}{1-\sum_{k=1}^{11} N'_{k,t}}\right).
\end{equation}

The terms in the coarse-grained partition (the right-hand-side of \eqref{eq:coarseMCT}) represent different components of the per capita growth rate: $\varepsilon^0 = r(\bar{E},\bar{C})$ is the baseline growth rate under mean environment and competition; $\varepsilon^E = \overline{r(E,\bar{C})} - \varepsilon^0$ represents the main effect of environmental variation; $\varepsilon^C = \overline{r(\bar{E},C)} - \varepsilon^0$ represents the main effect of competition variation; and $\varepsilon^{EC} = \overline{r(E,C)} - [\varepsilon^0 + \varepsilon^E + \varepsilon^C]$ captures the interaction between environment and competition. The environment-competition interaction term can be further decomposed as $\varepsilon^{EC} = \varepsilon^{(E\#C)} + \varepsilon^{(E \times C)}$, where $\varepsilon^{(E\#C)}$ is computed by shuffling the set of environmental variables to break any covariance between environment and competition.  The storage effect relies on the covariance between environment and competition, so the term representing this quantity is measured here as the total $EC$ interaction effect minus any non-covariance interaction: $\varepsilon^{(E \times C)} = \varepsilon^{EC} 
 - \varepsilon^{(E\#C)}$.

To assess how each $\epsilon$-component contributes to coexistence, we computed invader--resident comparisons. For example, the storage effect's contribution to the invasion growth rate is calculated as:
\begin{equation} \label{eq:inv_res_comparison_storage}
\text{(The storage effect)} \quad \Delta_i^{(E C)} = \varepsilon_{i|i}^{(E C)} - \frac{1}{S-1}\sum_{r\neq i} \varepsilon_{r|i}^{(E C)},
\end{equation}
where $i$ denotes the invader, $r$ indexes resident species, and $S$ is the number of species. Another important quantity is the comparative growth rate under mean conditions: 
\begin{equation} \label{eq:inv_res_comparison_fluctfree}
\text{(Fluctuation-free effect)} \quad \Delta_i^{0} = \varepsilon_{i|i}^{0} - \frac{1}{S-1}\sum_{r\neq i} \varepsilon_{r|i}^{0},
\end{equation}
The fluctuation-free effect $\Delta_i^{0}$ represents differences in performance when the MCT parameters ($E_j$ and $C_j$) are held at their mean values. This term serves as a null model against which fluctuation-dependent mechanisms can be compared, capturing species differences that would persist even in a constant environment.

\begin{figure}[H]
\centering
\includegraphics[scale = 1]{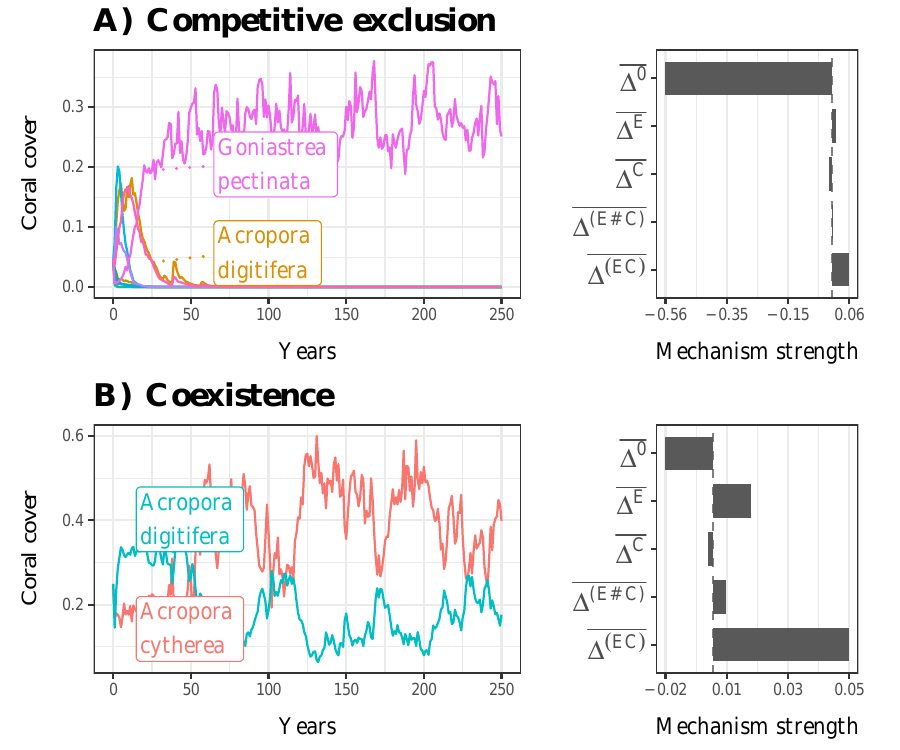}
\caption{Example communities with simulated time series and corresponding Modern Coexistence Theory partitions. In the case of competitive exclusion (\textbf{A}), the excluded species are hurt by mean conditions (i.e., large, negative $\overline{\Delta^0}$), which is not compensated by the storage effect (i.e., small, positive $\overline{\Delta^{(EC)}}$). In the case of coexistence (\textbf{B}), species suffer little from average conditions (i.e., small, negative $\overline{\Delta^0}$), and are stabilized by the storage effect (i.e., large, positive $\overline{\Delta^{(EC)}}$). Note the different ranges of the horizontal axes in the right-hand panels.}
\label{fig:TS_combined}
\end{figure}

We calculated analogous invader-resident comparisons for all components and computed community-average mechanisms by averaging across all possible invader-resident configurations for each parameter set and community-module type. These community averages (denoted $\overline{\Delta^0}$, $\overline{\Delta^E}$, $\overline{\Delta^C}$, $\overline{\Delta^{(E\#C)}}$, and $\overline{\Delta^{(EC)}}$) quantify the overall stabilizing strength of each process and form the basis of our results (e.g., Fig. \ref{fig:TS_combined}).
The $E_j$ and $C_t$ ``data'' are collected from simulations where single species have been perturbed to low density (details in Appendix \ref{Invasion analysis algorithm}).

For a more detailed understanding of storage effects, we performed a fine-grained decomposition where environmental parameters were separated into survival ($S$), growth ($G$), and fecundity ($F$) components, while competition parameters were divided into larval production ($L$) and occupied area ($A$). This yielded five biologically relevant storage effects: the covariance between survival fluctuations and larval densities ($\Delta_i^{(S \times L)}$); between survival fluctuations and occupied space ($\Delta_i^{(S \times A)}$; ``A'' for area); between growth fluctuations and larval densities ($\Delta_i^{(G \times L)}$); between growth fluctuations and occupied space ($\Delta_i^{(G \times A)}$); and between fecundity fluctuations and larval densities ($\Delta_i^{(F \times L)}$). These sub-storage effects allow us to identify which specific environmentally-dependent demographic processes interact with which sources of density dependence to promote species coexistence.

\section{Results}
\label{Results}

\begin{figure}[H]
\centering
\includegraphics[scale = 1]{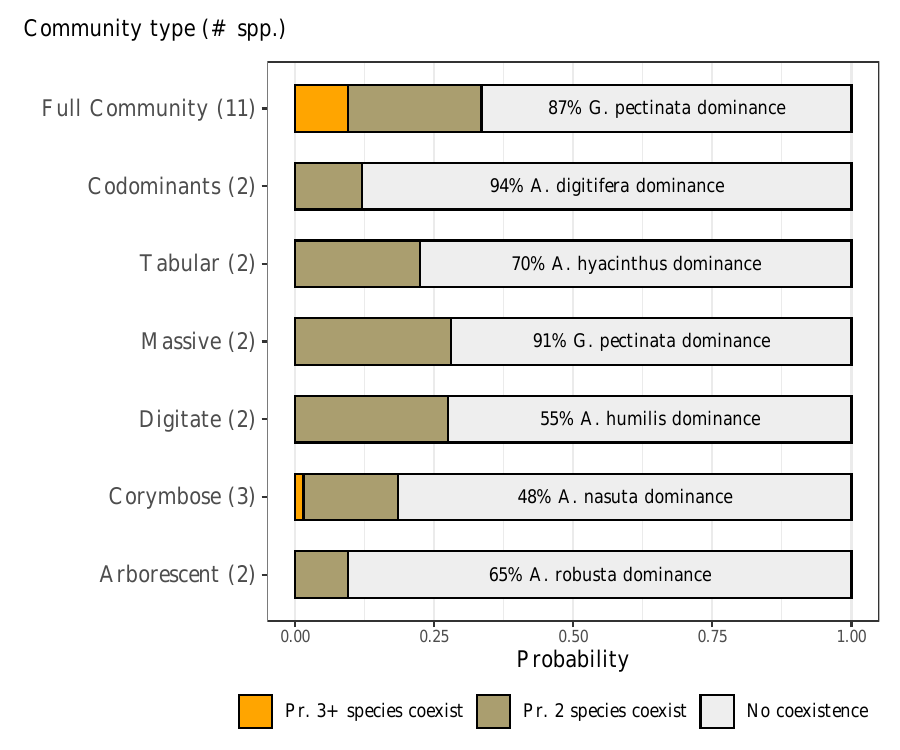}
\caption{The posterior probability of coexistence is low, both in the full community and across groups of species with similar morphologies. Colored bars represent the posterior probability of 2 or 3+ species coexisting. When only one species persists (i.e., ``No coexistence'') the in-figure text indicates which species it is and the proportion of ``No coexistence'' simulations in which it dominates. The ``Codominant'' community model contains only \textit{A. digitifera} and \textit{A. hyacinthus}, the two most common species at the study site.}
\label{fig:scenario_bars}
\end{figure}

Coral coexistence was uncommon in IPM simulations, with competitive exclusion being the dominant outcome across all community modules (Fig. \ref{fig:scenario_bars}). In the full-community module involving all 11 coral species, the posterior probability of two or more species coexisting was 38\%, but the probability of three or more species coexisting was only than 12\%. The probability of species pairs coexisting within morphological groups ranged from 10\% to 30\%. 

In cases of competitive exclusion within the full community, \textit{Goniastrea pectinata} emerged as the dominant species, winning in 88\% of simulations (Fig. \ref{fig:scenario_bars}). This result is unexpected, as \textit{G. pectinata} was neither the most abundant coral species on the reef nor the most abundant within its ``massive'' morphological group \citep{madin2023six}. Analysis of demographic rates (Figures \ref{fig:surv_fit} \& \ref{fig:growth_fit}) shows that both \textit{Goniastrea} species exhibit unusually high juvenile survival rates, but only \textit{G. pectinata} demonstrates high juvenile growth. Two potential explanations could account for why the simulation results differ from real-world observations of \textit{G. pectinata} abundance. First, the IPM may not incorporate key factors such as spatial niche partitioning. Second, \textit{Acropora} cover declined disproportionately during the study period \citep{madin2018cumulative}, likely as a result of temperature stress disproportionately affecting \textit{Acropora spp.} \citep{loya2001coral}, such that the estimated demographic rates may not reflect long-term, historical demographic rates. However, it is important to note that our primary finding --- the rarity of coexistence via the storage effect --- remains true regardless of \textit{G. pectinata}'s virtual dominance, since coexistence was uncommon across all community modules.

\begin{figure}[H]
\centering
\includegraphics[scale = 1]{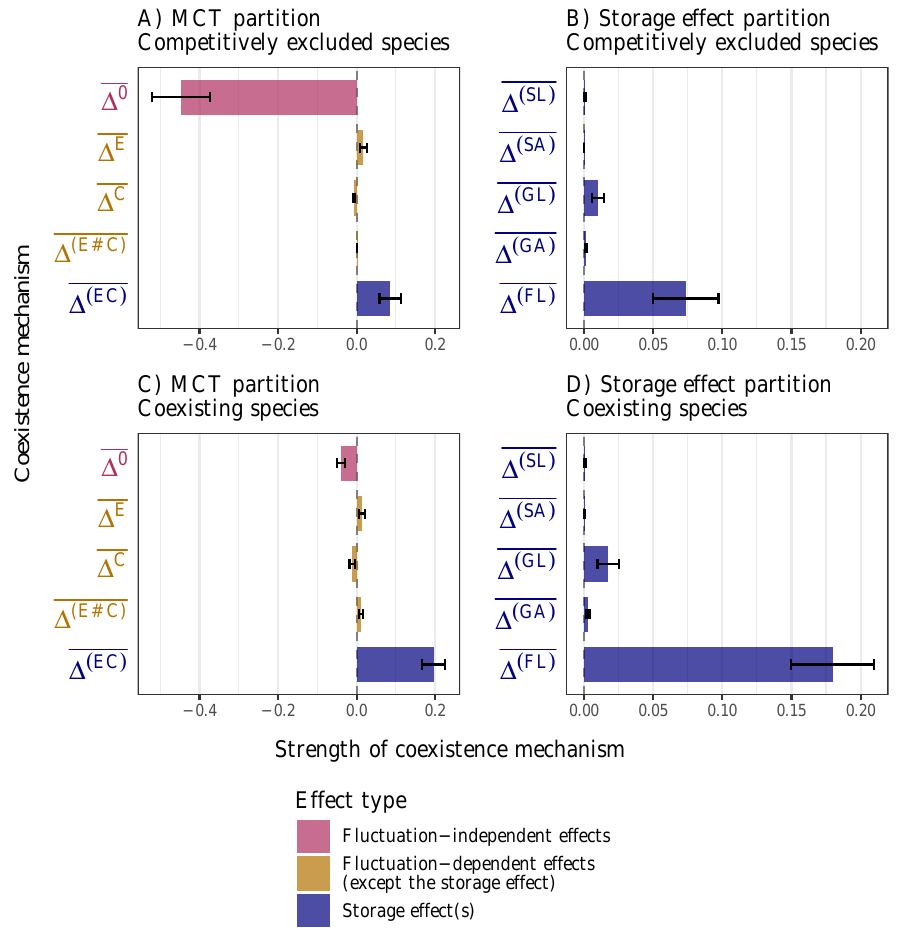}
\caption{The Modern Coexistence Theory partition reveals two key findings about the storage effect. First, coexistence occurs when model parameters produce a weak storage effect ($\overline{\Delta^{(EC)}}$ in both panels A and C) while minimizing fitness differences (shown by much less negative $\overline{\Delta^0}$ in panel C compared to A). Among the five storage effects examined, only fecundity fluctuations produce a substantial effect (evidenced by large $\overline{\Delta^{(FL)}}$ in panels B and D), which is consistent with the ``classic storage effect'' found in simple models of fishes and annual plants. Panels show coexistence mechanism strength across competitively excluded species (top row), coexisting species (bottom row), the classical MCT partition (left column), and the process-based MCT partition showing sub-storage effects (right column). Bars and error bars respectively show the Mean and SE across both community modules (e.g., all species, just tabular species) of the posterior means.}
\label{fig:MCT_partition_combined}
\end{figure}

While the storage effect promoted species coexistence, it generally wasn't strong enough to overcome mean fitness differences between coral species. Our simulations revealed that fluctuation-free effects were the primary drivers of outcomes, typically being five times larger than storage effects (compare ($\overline{\Delta^0}$ and $\overline{\Delta^{(EC)}}$ in Fig. \ref{fig:MCT_partition_combined}, panel A). When comparing cases of species coexistence versus competitive exclusion, we found that the storage effect was only modestly greater in coexistence than exclusion scenarios. More significantly, the fluctuation-free effects shifted from strongly negative values in exclusion cases to near-zero values in coexistence cases. This disparity suggests that when species did coexist, they did so because the particular demographic parameters equalized species' fitnesses, rather than conferring an unusually strong stabilizing effect via temporal niche partitioning.

Among the five process-based storage effects we quantified, the fecundity-to-larvae storage effect was the largest, followed by the growth-to-larvae storage effect (see $\overline{\Delta^{(FL)}}$ and $\overline{\Delta^{(GL)}}$ in Fig. \ref{fig:MCT_partition_combined}, panels B and D).  This pattern was consistent across most community modules, with a few exceptions; specifically, the growth-to-larvae storage effects were negligible in the codominant, digitate, and massive modules. Notably, the fecundity-to-larvae storage effect is the classical storage effect, first introduced in the lottery model of coral reef fish coexistence \citep{chesson1981environmentalST}.

\begin{figure}[H]
\centering
\includegraphics[scale = 1]{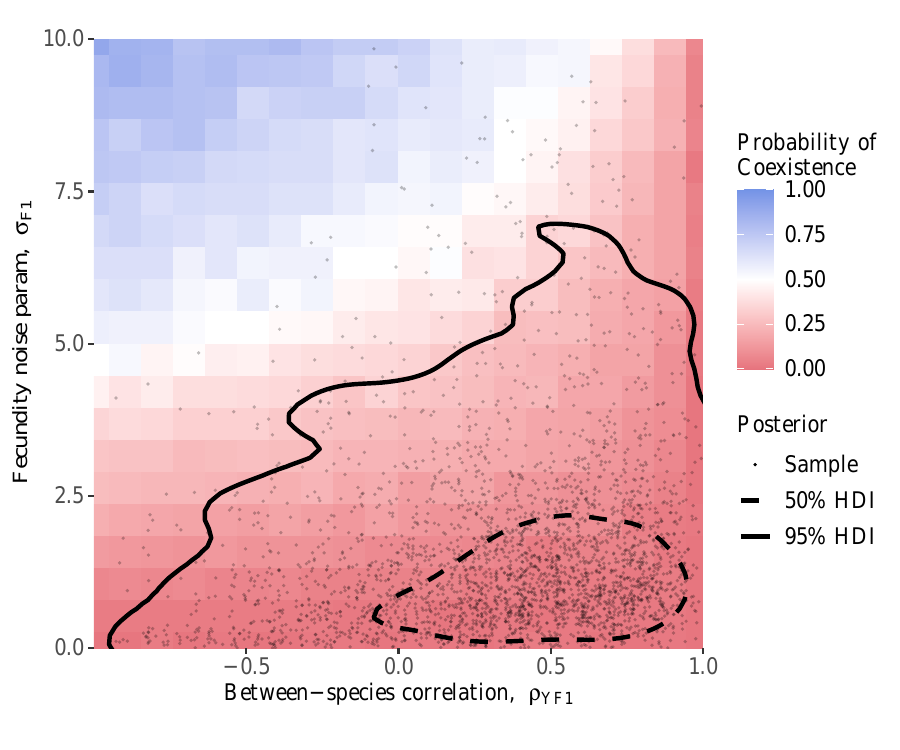}
\caption{Coexistence via the storage effect occurs when species have strong environmental niche differences (low $\rho_{F1}$) combined with high environmental variation (high $\sigma_{j,YF1}$), but our fitted model indicates that species fall far outside this regime. The heatmap color shows the posterior probability of coexistence between two empirically-codominant species, \textit{A. digitifera} and \textit{A. hyacinthus}, as a function of two parameters: the between-species correlation ($\rho_{F1}$) and the scale of environmentally-driven variation ($\sigma_{j,YF1}$) in coral fecundity.  Posterior probabilities of coexistence were calculated from 100 model simulations with fixed $\rho_{F1}$ and $\sigma_{j,YF1}$. Marginal distributions of these parameters are shown by posterior samples and high-density intervals (HDI).}
\label{fig:high_var_tiles}
\end{figure}

Sensitivity analyses revealed that coexistence could only emerge under conditions of nearly equal fluctation-free fitnesses (noted earlier), greatly exaggerated environmental variability, or unusually low interspecific correlations. We manipulated colony fecundity by adjusting the probability of colonies being reproductive/gravid, systematically varying both the scale of environmental variability and interspecific correlations. The results showed that high-probability stable coexistence could only emerge under extreme conditions: either environmental fluctuations $\sim 10 \times$ greater than observed values, or a combination of $\sim 5 \times$ fluctuations with strong negative correlations between species' environmental responses (Fig.~\ref{fig:high_var_tiles}). When we alternatively manipulated colony fecundity by adjusting variation in the number of eggs produced by gravid colonies, the coexistence regime was even farther away from empirical estimates (Fig. \ref{fig:high_var_tiles_F2}). 

Our main finding --- that the storage effect makes at most a negligible small contribution to species coexistence --- is robust to different modeling approaches and parameterizations (Appendix \ref{Robustness analysis}). When we excluded wave-disturbance mortality to assess coexistence in non-reef-crest environments, coexistence probabilities decreased slightly compared to the baseline model. Coexistence probabilities decreased further when we implemented an alternative growth model (proposed by \citealp{madin2020partitioning}). Incorporating diffuse correlations between species, to account for uncertain interspecific demographic correlations more conservatively, also consistently reduced coexistence likelihood. Finally, three scenarios considering alternative recruitment dynamics produced coexistence probabilities that were either lower than or comparable to our baseline results.

\section{Discussion}
\label{Discussion}

The storage effect (i.e., temporal niche partitioning) appears to play only a minor role in maintaining coral species diversity in our western Pacific study system. This finding is especially notable because coral communities possess the key characteristics thought necessary for temporal niche partitioning (large recruitment fluctuations, long-lived adult colonies, and a propagule stage susceptible to density-dependent processes) while showing little evidence of alternative coexistence mechanisms such as resource or natural enemy partitioning (see \textit{Introduction} for evidence). Finding a weak storage effect in a system where it should theoretically be strong suggests a broader pattern: the storage effect may be generally weak across ecosystems. While we cannot generalize with absolute certainty, our results align with an emerging trend in the literature showing that when explicitly quantified, the storage effect typically plays only a minor role compared to other coexistence mechanisms  (\citealp[Table 6]{Ellner2019}; \citealp{hallett2019rainfall, armitage2019negative, armitage2020coexistence, zepeda2019fluctuation, letten2019mechanistic, stump2023reexamining}).

Among the five storage effects investigated, the one based on fluctuations in fecundity (i.e., $\overline{\Delta^{(FL)}}$) had the strongest influence. This result validates the historical focus on fluctuations in fecundity and propagule establishment (e.g., \citealp{warner1985coexistence, caceres1997temporal, hallett2019rainfall}), processes that directly influence recruitment. While the storage effect could theoretically arise from variation in any demographic rate, our findings suggest that future studies of the storage effect should continue to prioritize fecundity and propagule establishment.

Storage effects associated with survival and growth were comparatively weak. In Appendix \ref{Decomposing Storage Effects}, we used mathematical analysis and simulations to understand why, decomposing these storage effects into their fundamental components: species-specific environmental responses, an $EC$ covariance, and non-additivity (see \textit{Introduction}). The growth-based storage effect is weak primarily because it lacks a positive $EC$ covariance. This absence likely occurs because competition depends on colony size distributions, which are not substantially altered by short-term favorable growth conditions. The survival-based storage effect is weak for a different reason: at equilibrium, species' growth rates do not respond strongly to fluctuations in wave disturbances. This is not to say that wave disturbances are unimportant for coral dynamics. Rather, for species that can coexist, deviations from mean disturbance levels do not lead to large fluctuations in per capita growth rates.

We considered whether our finding of minimal storage effects might be an artifact of underestimating environmental fluctuations in coral communities. While coral coexistence might theoretically depend on extreme fluctuations occurring at longer timescales (decades or centuries) than our 5-year dataset captures, several lines of evidence argue against this alternative explanation. First, environmental fluctuations would need to be at least $10 \times$ larger than current levels to make coexistence probable through the storage effect alone (Figures \ref{fig:high_var_tiles} \& \ref{fig:high_var_tiles_F2}). Second, fecundity fluctuations in our model are large in absolute terms, with average reproductive probabilities ranging from near zero to nearly 100\% (Fig. \ref{fig:fecundity_var_TS}), a dynamic previously observed on the Great Barrier Reef \citep{hughes2000supply}. Finally, our modeled fecundity fluctuations are larger than or comparable to those reported across six different coral communities (Appendix \ref{Fecundity fluctuations in the literature}).

While parameter uncertainty in our model could potentially mask a more significant storage effect, this alternative explanation faces several challenges. We could not directly estimate maximum recruit density parameters ($\beta_{j,R}$) from data, so we selected values consistent with the broad range of coral cover observed in natural communities. This procedure led to substantial parameter uncertainty, which could hypothetically mask a situation where species' demographic parameters generate nearly equal fitness levels, thereby enhancing the storage effect's importance. However, this scenario is difficult to verify because settlers are difficult to identify to species-level, and because the scenario entails fine-tuning of demographic rates --- either through local adaptation or selection from a regional species pool --- that seems unlikely to withstand invasive species or secular trends in the environment. Given these limitations, we should explore additional coexistence-promoting mechanisms to explain apparent coexistence in the face of a weak storage effect.

Many coral species exhibit broad habitat preferences along gradients such as depth, wave exposure, temperature, and sedimentation tolerance \citep{anthony2004environmental, arias2008scaling, osborne2011disturbance, mellin2020representation}. These differences reflect spatial (environmental) niche partitioning, known within modern coexistence theory as the \textit{spatial storage effect} or \textit{fitness-density covariance} \citep{chesson2000general}. However, large-scale habitat differentiation alone seems insufficient to explain the exceptionally high within-habitat (alpha) diversity found in coral communities. Coral species frequently have broad, overlapping depth distributions \citep{fricke1985depth, carpenter2022light}, rather than the distinct zonation that would intuitively support hundreds of coexisting species. An alternative explanation is that corals achieve coexistence through microhabitat partitioning. Coral larvae demonstrate preferences for centimeter-scale environmental factors including flow patterns, sediment loads, surface roughness, light intensity, and specific types of crustose coralline algae (CCA) for settlement (\citealp{bak1979distribution, babcock1996coral, edmunds2004effects, vermeij2008density, mason2011coral, strader2015differential, ricardo2017settlement, martinez2023three}, \citealp{ritson2016patterns}; \citealp{jorissen2021coral} but see \citealp{harrington2004recognition}). 

Species-specific natural enemies could promote coexistence if their effects become stronger as conspecific densities become greater (i.e., the Janzen-Connell effect; \citealp{janzen1970herbivores}). In tropical plants, this phenomenon occurs largely when fungal pathogens of adults are shed into the nearby soil, causing  propagule establishment to be inhibited more in the vicinity of conspecific than heterospecific adults \citep{spear2021host}. In corals, pathogenic bacteria from adult colonies have been shown to harm settlers in some studies \citep{edmunds2004effects, marhaver2013janzen, sims2021janzen}, but other studies have found no evidence of negative conspecific effects \citep{gibbs2015spatial}, and crucially, no experiments have tested larvae with both conspecific and heterospecific adults in a reciprocal design. 

% Moreover, even if present, whether and by how much such natural enemy effects are coexistence-promoting is itself context-dependent (cite plant papers).

Lastly, coexistence may occur via source-sink dynamics, with different species being advantaged in different habitat patches, and local coexistence occurring via the dispersal and establishment of sub-populations in environments from which they would be competitively excluded in the absence of such dispersal \citep{mouquet2003community}. Differential performance of different species in different patches are one mechanism by which such a phenomenon could occur, but another is spatial asymmetries in the dispersal of larvae. At the metacommunity scale, species-specific spawning times, larval survival, and competency timing \citep{connolly2010estimating} lead to different larval transport patterns that concentrate the larvae of different species on different reefs \citep{berkley2010turbulent, salomon2010effects, figueiredo2012dispersal}. In this fashion, average competition may be higher among conspecifics than heterospecifics \citep{salomon2010effects}, producing a spatial storage effect \textit{sensu} \citep{chesson2000general}.

More broadly, our results contribute to accumulating evidence that fluctuations-mediated coexistence mechanisms play a minor role in biodiversity maintenance (e.g., \citealp{hallett2019rainfall, zepeda2019fluctuation, stump2023reexamining}). Previous empirical studies of the storage effect have typically focused on fecundity-based mechanisms without exploring robustness to parameter uncertainty or alternative model structures. By demonstrating that storage effects remain weak across a broad region of parameter space, under multiple model formulations, and with fluctuation magnitudes comparable to those reported in the literature, we greatly strengthen the case against this coexistence mechanism. This result, combined with the fact that coral communities possess many characteristics that should favor strong storage effects, suggests that the storage effect is weak in most communities. For coral communities specifically, we suspect that coexistence relies on alternative mechanisms involving spatial heterogeneity, larval behaviors, and dispersal dynamics.

The apparent weakness of the storage effect suggests that classical niche differentiation deserves renewed attention. Several studies have documented species in high-diversity systems partitioning a single niche axes, such as light utilization in tropical trees and phytoplankton \citep{kitajimar2008functional, schwaderer2011eco}, while other studies reveal high-dimensional niche spaces where species specialize on unique combinations of limiting factors \citep{mcmahon2011high, ingram2022hierarchical, kalyuzhny2023pervasive}. Particularly intriguing are recent discoveries of cryptic niche differentiation, such as fine-scale dietary partitioning in termites and large herbivores \citep{schyra2019cryptic, pansu2022generality}, or depth partitioning among cryptic coral species \citep{johnston2022niche}. However, while many studies have identified niche differences, few have explicitly quantified the relative strength of different coexistence mechanisms to determine which niche dimensions are most important \citep{adler2013trait}, and we know of no attempts to make such an assessment in corals or other coral reef taxa. Such quantitative comparisons, enabled by frameworks like modern coexistence theory, are essential for understanding biodiversity maintenance.

\section{Acknowledgements}  \label{Acknowledgements} 

We would like to thank Carrie Sims and Alfonso Ruiz for helpful conversations. This work is supported by a Rubinoff Big Bet grant to S.R.C. from the Smithsonian Tropical Research Institute.

\begin{appendices}
\counterwithin{figure}{section}
\counterwithin{table}{section}
\counterwithin{equation}{section}

\section{Model description} \label{Model description}

The model description follows the sequence of events that occur within each time step: mortality (Appendices \ref{Wave action survival}--\ref{Background mortality}, growth (\ref{Colony growth}), spawning (\ref{Fecundity}), and recruitment (\ref{Recruitment}). Detailed methods for parameter estimation and model diagnostics are in Appendix \ref{Model-fitting details}, and simulation model implementation details are in Appendix \ref{IPM implementation}. Table \ref{tab:symbols} defines model parameters and other relevant symbols.

\subsection{Wave-disturbance mortality} \label{Wave action survival}

We used established hydrodynamic and biomechanical models to predict coral colony dislodgement from wave action \citep{madin2006scaling, madin2006ecological, madin2012calcification, alvarez2023disturbance}. Dislodgement risk increases with water velocity, colony size, and top-heaviness. Since these broken-off colonies rarely survive \citep{smith1999experimental}, we treat dislodgement as fatal. 

The wave-disturbance sub-model works in three main steps. First, we generate annual maximum wind velocities $v_t$ (in $\text{m},\text{s}^{-1}$) from a gamma distribution: $v \sim \text{Gamma}(\alpha_W, \beta_W)$, with $\alpha_W = 2.18$ and $\beta_W = 0.35$. Parameter values come from \citet{alvarez2023disturbance}, who analyzed historical wind data from Australian Bureau of Meteorology measurements near Lizard Island.

Second, annual wind velocities are converted to water velocities $u_t$ (in $\text{m},\text{s}^{-1}$) with the saturating function,
\begin{equation} \label{eq:wind_water}
u = a(1 - e^{-bv}), 
\end{equation}
with $a = 5.10$ and $b = 0.04$. \citet{madin2006ecological} originally estimated water velocities using the WGEN wave generation model \citep{black1990reef, black1992natural}, with predictions validated against field measurements at the study site. \citet{madin2012calcification} showed that saturating functions provide a good fit to the relationship between wind and water velocities; fetch and water depth naturally limit wave development. Then, \citet{alvarez2023disturbance} used this dataset of wind velocity measurements and water velocity estimates to fit \eqref{eq:wind_water} via least squares.

Third, the dislodgment mechanical threshold (DMT) and the colony shape factor (CSF) together determine disturbance-induced mortality \citep{madin2006scaling}. The year-specific DMT measures disturbance intensity, and is defined as
\begin{equation}
\text{DMT}_t = \frac{\sigma_s}{u_t^2\rho_w},
\end{equation}
where $\rho_w$ is seawater density (1025 kg m$^{-3}$), and $\sigma_s$ is the substrate tensile strength (0.2 MNm$^{-2}$). These constants are based on \textit{in situ} measurements \citep{madin2005mechanical}.

The colony shape factor (CSF) is a measure of coral vulnerability. For a particular coral colony, CSF is defined as:
\begin{equation}
\text{CSF} = \frac{16}{d_{\parallel}d_{\perp}} \int_0^h w(y)dy.
\end{equation}
Here, $d_{\parallel}$ and $d_{\perp}$ represent the perpendicular basal attachment widths, $w(y)$ denotes the colony width projected at height $y$, and $h$ is the maximum colony height above the substrate.

\begin{figure}[H]
\centering
\includegraphics[scale = 1]{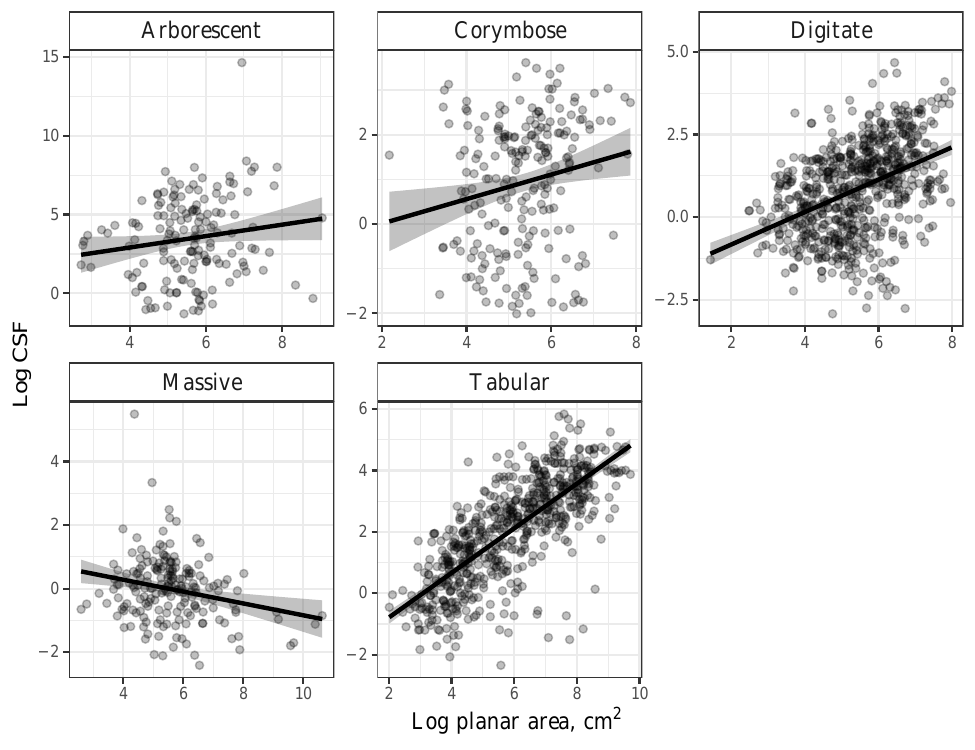}
\caption{Log CSF as a function of log planar area, stratified by colony morphology. Lines and grey ribbons respectively show the posterior mean and 95\% credible intervals of $\beta_{m[i],0C} + \beta_{m[i],1C} \, x_i$.}
\label{fig:CSF_by_area_fit}
\end{figure}

We cannot directly measure CSF for simulated colonies, because our model only tracks colony area, not 3D shape. However, CSF can be estimated based on colony area and colony morphology. The Lizard Island dataset contains 11 species across 5 morphology categories: arborescent, tabular, corymbose, digitate, and massive. Using 1846 observations of CSF/planar area from \citet{madin2014mechanical}, we parameterized the model
\begin{equation}
\log(\text{CSF}_i) \sim \text{Normal}\left(\beta_{m[i],0C} + \beta_{m[i],1C} \, x_i, \sigma_{m[i],C} \right),
\end{equation}
where $i$ indexes individual observations and $m[i]$ indicates the morphology group. The independent variable $x_i$ is the logarithm of colony planar area (in $\text{cm}^2$). The parameters $\beta_{m,0C}$ and $\beta_{m,1C}$ represent the morphology-specific intercept and slope terms, while $\sigma_{m,C}$ represents the morphology-specific residual error. The model fit is visualized in Figure \ref{fig:CSF_by_area_fit}.

For model simulations, we estimate CSF using the regression mean. For species $j$ with morphology $m$ and log planar area $x$:
\begin{equation}
    \widehat{\text{CSF}}_j(x) = \exp\left(\beta_{m[i],0C} + \beta_{m[i],1C} \, x_i, \right).
\end{equation}
This simplified approach doesn't account for CSF variation among same-sized colonies. We explored a more complex model allowing variable CSF for 1-year recruits, followed by deterministic CSF evolution based on colony growth. However, this alternative approach is computationally costly due to a high-dimensional state space (e.g., 50 bins for planar area $\times$ 50 CSF bins produces a 2500-element state for each species). Since preliminary simulations produced qualitatively identical results --- the storage effect doesn't engender many-species coexistence --- we adopted the simpler approach, following \citet{alvarez2023disturbance}.

For simulations of the IPM, the probability of colony mortality due to wave disturbance, $D_{j,t}(x)$, is given by
\begin{equation} \label{eq:D}
D_{j,t}(x) = \begin{cases}
1 & \text{if } \widehat{\text{CSF}}_j(x) < \text{DMT}_t. \\
0 & \text{otherwise}
\end{cases}
\end{equation}

The wave-disturbance sub-model differs from other sub-models (e.g., growth, fecundity) because it utilizes parameter values from previous research. Previous studies provide point estimates based on summary statistics or maximum likelihood estimation, so the parameters $\alpha_W$, $\beta_W$, $a$, $b$, $\rho_W$, and $\sigma_t$ are treated as fixed quantities. The parameters $\beta_{m,0C}$, $\beta_{m,1C}$, and $\sigma_{m,C}$ are fit to data using Bayesian methods, and thus vary across simulations.

\subsection{Background mortality} \label{Background mortality}

\begin{figure}[H]
\centering
\includegraphics[scale = 1]{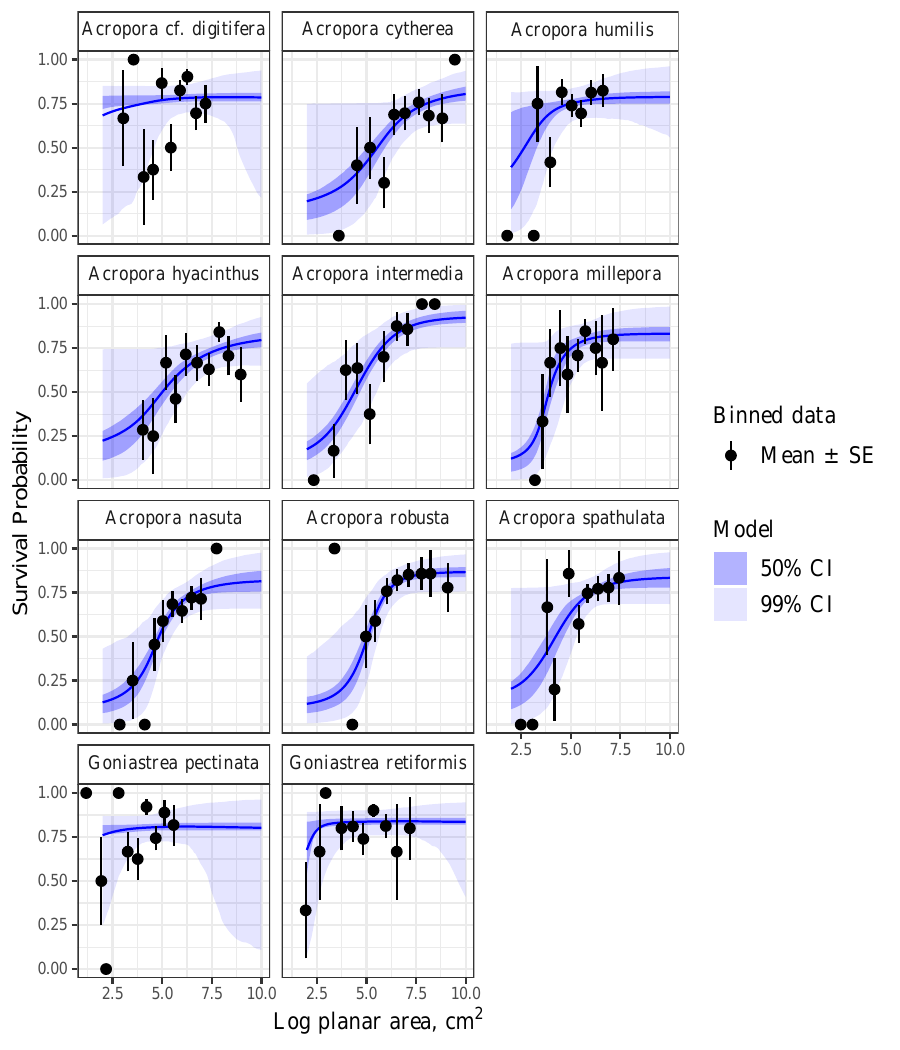}
\caption{Comparison between observed and predicted survival rates across colony size. Data points show the mean observed survival rates (± standard error) calculated from 10 equally sized bins spanning each species' size range. The blue line represents the posterior mean, with the shaded regions showing credible intervals (CI).}
\label{fig:surv_fit}
\end{figure}

Background mortality encompasses all causes of coral colony death except those resulting from wave disturbances. Here, we model the annual probability of survival as a logistic function of colony size (Fig. \ref{fig:surv_fit}). Importantly, we do not incorporate temporal variation in these background mortality rates. Since environmental stochasticity already affects population dynamics through the wave-disturbance mortality sub-model, adding year effects to background mortality could ``double count'' the effect of environmental stochasticity.

The statistical model uses 1645 observations from 2009--2015. Let $\text{surv}_{i}$ be a binary variable indicating whether colony $i$ survived ($\text{surv}_{i} = 1$) or died ($\text{surv}_{i} = 0$). Survival is then modeled as 
\begin{equation} \label{eq:base_surv1}
\text{surv}_{i} \sim \text{Bernoulli}\left( \text{logit}^{-1}(\beta_{j[i],0S} + \beta_{j[i],1S} \cdot x_i)\right)
\end{equation}
where $x_i$ is the planar colony area in $\text{cm}^2$. The parameters $\beta_{j[i],0S}$ and $\beta_{j[i],1S}$ the species-specific intercept and slope parameters. The subscript $j$ is the species index.

For simulations of the IPM, the mortality fraction is given by
\begin{equation} \label{eq:M}
M_j(x) = 1 -  \text{logit}^{-1}(\beta_{j,0S} + \beta_{j,1S} \cdot x).
\end{equation}

Combining this with wave-disturbance morality (sub-model \ref{Wave action survival}), the total annual survival probability can be written as
\begin{equation}
S_{j,t}(x) = \left(1- M_j(x)\right)\left(1- D_{j,t}(x)\right).
\end{equation}

\subsection{Colony growth} \label{Colony growth}

Our growth model incorporated both species identity and colony size as key factors influencing growth rates (Fig. \ref{fig:surv_fit}). The statistical model uses 1,170 growth observations collected between 2009-2013. Environmentally-driven fluctuations in growth were captured with year effects. A Box-Cox transformation was applied to satisfy model assumptions of linearity and homoscedasticity. A Student's $t$ error distribution was employed to accommodate the wide range of observed growth rates.

A single observation $i$ includes a colony's log planar area (in $\text{cm}^2$) in the focal year, denoted $x_i$, and the log planar area in the next year (assuming survival), denoted $x'_i$. The colony growth rate is the quotient: 
\begin{equation}
G_{i} = \frac{\exp(x'_i)}{\exp(x_i)}.
\end{equation}
Note that $x_{i}$ is on the logarithmic scale, so $\exp(x_{i})$ gives the raw area. To improve the statistical properties of our model, we applied a Box-Cox transformation:
\begin{equation}
g = \begin{cases}
\log(G) & \text{if } \lambda = 0 \\
\frac{G^\lambda - 1}{\lambda} & \text{otherwise},
\end{cases}
\end{equation}
where $\lambda$ is estimated from the data using maximum likelihood. We obtained $\lambda = 0.719$ based on all data (pooling years and species). The Box-Cox transformation helps normalize the distribution of colony growth rates and stabilize the variance.

The transformed growth rate is modeled using a Student's t-distribution:
\begin{equation}
g_{i} \sim \text{Student-}t(\nu_{j[i]}, \mu_{j[i],t[i],G}, \sigma_{j[i]}),
\end{equation}
where $\nu_j$ is the species-specific degrees of freedom parameter, $\sigma_j$ is the species-specific residual standard deviation. Initially, we used a normal error distribution, but Q-Q plots revealed heavy tails. The mean $\mu_{j,t,G}$ depends on colony size, species identity, and year:
\begin{equation} \label{eq:mu_G}
\mu_{j,t,G} = \beta_{j[,0G} + \beta_{j,1G} \cdot x_{i} + \eta_{j,t,G}. 
\end{equation}
The $\eta_{j,t,G}$ are species-specific year effects, here modeled as random intercepts in a non-centered parameterization (i.e., the mean, $\beta_{j,0G}$, is separated out). The year effects have a hierarchical structure: 
\begin{equation}
\eta_{j,t,G} \sim \text{Normal}(0, \sigma_{j,YG}),
\end{equation}
where the hyperparameter $\sigma_{j,YG}$ determines the level of variation.

For each sample of the posterior distribution, and for each species pair $(j,k)$, we calculate the Pearson correlation coefficient $\rho_{j,k,G}$ of their year effects, producing the correlation matrix $\mathbf{R}_G$. This matrix captures how similarly different species grow in response to environmental fluctuations.

To generate random year effects for IPM simulations, we first construct a covariance matrix $\mathbf{\Sigma}_G$ from the correlation matrix and species-specific standard deviations:
\begin{equation}
\Sigma_{j,k,G} = \sigma_{j,YG} \, \sigma_{k,YG} \, \rho_{j,k,G}.
\end{equation}
Then, for each simulated year $t$, we sample the vector of species-specific year effects $\boldsymbol{\eta}_{t,G} = (\eta_{1,t,G}, \ldots, \eta_{11,t,G})$ from a multivariate normal distribution:
\begin{equation}
\boldsymbol{\eta}_{t,G} \sim \text{MVN}(\mathbf{0}, \mathbf{\Sigma}_G).
\end{equation}
This sampling approach attempts to preserve the correlation structure between species' responses to the environment, without explicitly modeling the correlations. We do, however, explicitly model the correlations (i.e., treat them as parameters) in the robustness analysis (Appendix \ref{Robustness analysis}). 

For simulation of the IPM, we compute growth transitions between size bins through numerical integration. For a colony of species $j$ and log planar area $x$, we generate the vector of year effects and calculate the mean growth factor (via \eqref{eq:mu_G}). Then, we evaluate the Student's $t$-density --- using species-specific $\nu_j$ and $\sigma_{j,G}$ --- across all possible size transitions. To improve computational efficiency, we pre-compute the Box-Cox transformation matrix for all possible size transitions. 

For simulations of the IPM, the continuous state space of colony sizes is discretized, growth variables ($g$) are computed for each possible size transition, and transition probabilities are evaluated with the Student's $t$-distribution. To improve computational efficiency, we pre-compute the Box-Cox transformation matrix for all possible size transitions.

\begin{figure}[H]
\centering
\includegraphics[scale = 1]{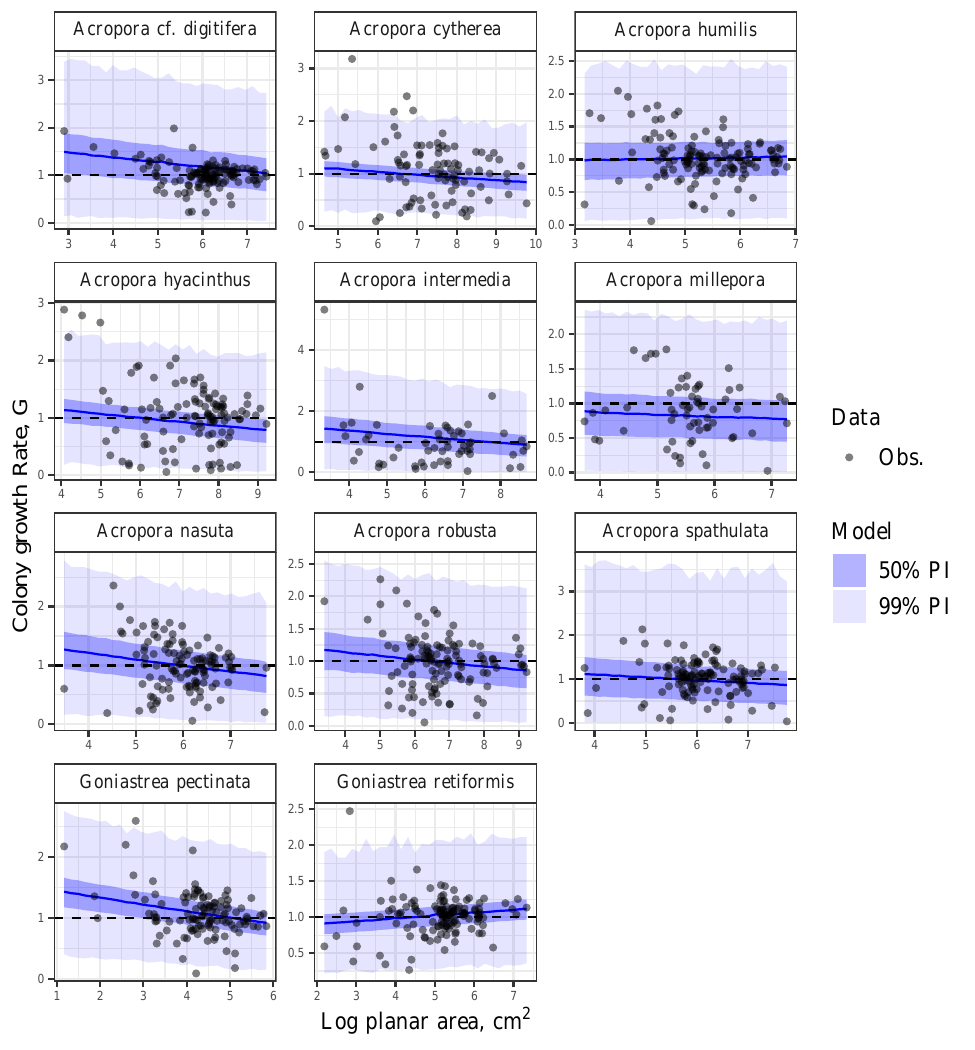}
\caption{Comparison between observed and predicted growth rates across colony size. Each data point represents a single colony's growth in a particular year. The blue line represents the posterior mean, with the shaded regions showing predictive intervals (PI).}
\label{fig:growth_fit}
\end{figure}

\subsection{Coral fecundity} \label{Fecundity}

\begin{figure}[H]
\centering
\includegraphics[scale = 1]{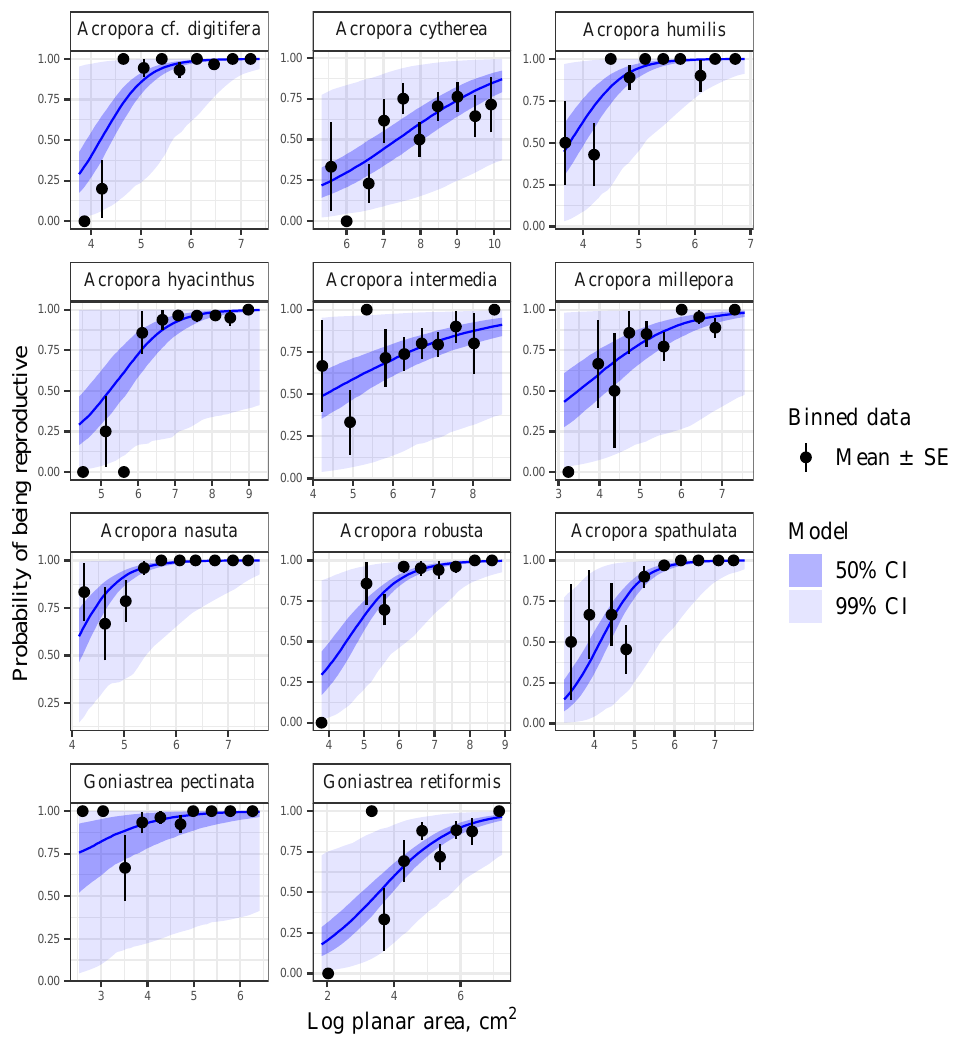}
\caption{Comparison between observed and predicted reproductive probability across colony size. Data points show the proportion of reproductive colonies (± standard error) calculated from 10 equally-sized bins spanning each species' size range. The blue line represents the posterior mean, with the shaded regions showing credible intervals (CI).}
\label{fig:fecundity1_fit}
\end{figure}

\begin{figure}[H]
\centering
\includegraphics[scale = 1]{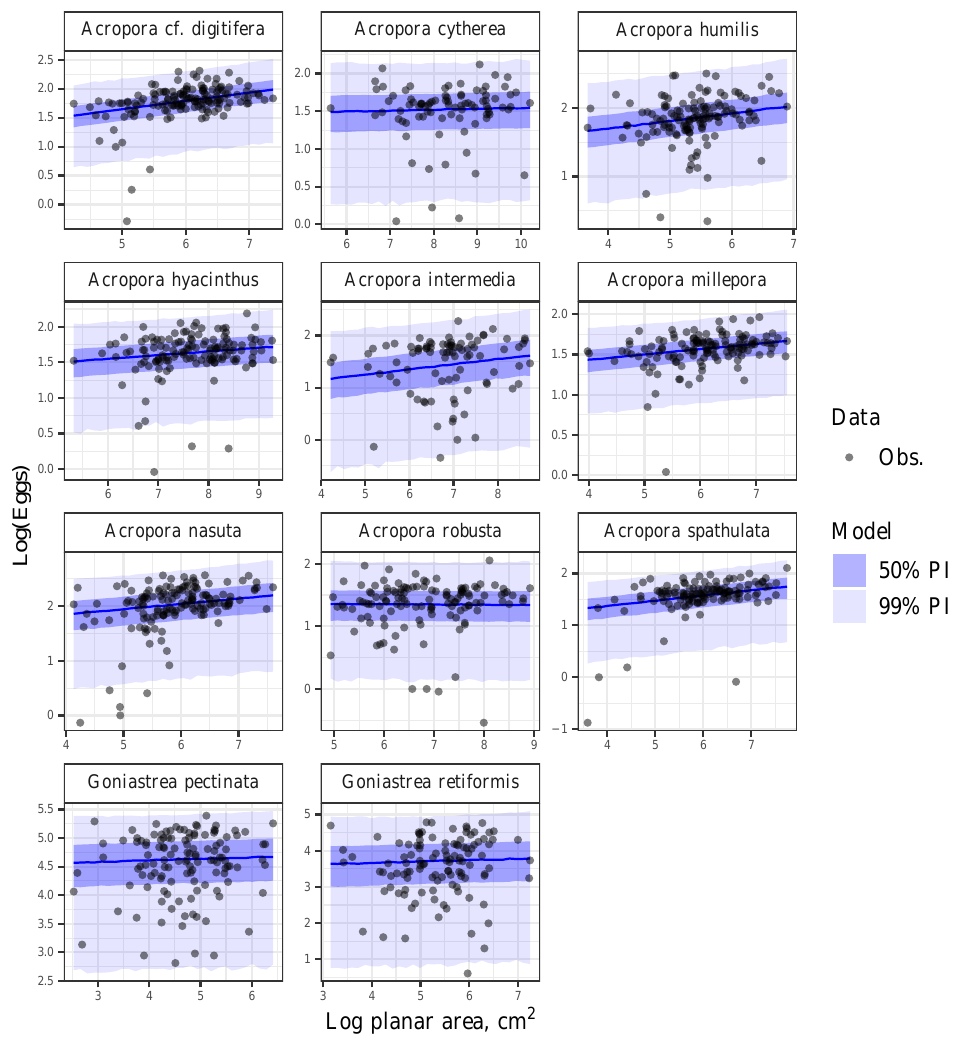}
\caption{Comparison between observed and predicted eggs per polyp (conditioned on a colony being reproductive) across colony size. Each data point represents a single colony's egg production in a particular year. The blue line represents the posterior mean, with the shaded regions showing predictive intervals (PI).}
\label{fig:fecundity2_fit}
\end{figure}

We modeled coral fecundity using a log-linear hurdle model that accounts for both the probability of a colony being reproductive (Fig. \ref{fig:fecundity1_fit}), and if reproductive, the number of eggs produced per polyp (Fig. \ref{fig:fecundity2_fit}). In cases where multiple samples were taken from each colony (within a year), we computed the mean to obtain average eggs per polyp at the colony level. We used data from the years 2009-2014 ($n=1559$ colony-level observations). 

For a single colony $i$ with log planar area $x$ (in $\text{cm}^2$), let $\text{reproductive}_{i}$ be a binary variable indicating whether the colony produced eggs ($\text{reproductive}_{i} = 1$) or not ($\text{reproductive}_{i} = 0$). The hurdle part of the fecundity model is
\begin{equation} \label{eq:reproductive1}
\text{reproductive}_{i} \sim \text{Bernoulli}(\text{logit}^{-1}(\gamma_{j[i],0F} + \gamma_{j[i],1F} \cdot x_{i,t} + \eta_{j[i],t[i],F1})),
\end{equation}
where $\gamma_{j,0F}$ and $\gamma_{j,1F}$ are species-specific intercept and slope parameters respectively; and $\eta_{j,t,F1}$ are species-specific year effects. 

For colonies that are reproductive ($\text{reproductive}_{i} = 1$), let $\text{eggs}_{i}$ denote the mean number of eggs produced per polyp. We utilized a skew-normal distribution based on graphical evidence:
\begin{equation} \label{eq:log_eggs1}
\log(\text{eggs}_{i,t}) \sim \text{SN}(\xi_{j[i],t[i]}, \omega_{j[i]}, \alpha_{j[i]}),
\end{equation}
where $\text{SN}(\xi, \omega, \alpha)$ denotes the skew-normal distribution with location parameter $\xi$, scale parameter $\omega$, and shape parameter $\alpha$. The location parameter depends on colony size, species identity, and year:
\begin{equation}
\xi_{j[i],t[i]} = \beta_{j[i],0F} + \beta_{j[i],1F} \cdot x_{i} + \eta_{j[i],t[i],F2},
\end{equation}
Here, $\eta_{j,t,F2}$ are species-specific year effects for the egg count model. The scale and shape parameters determine the degree and direction of skewness in egg production.

The year effects for both parts of the hurdle model have hierarchical structures:
\begin{align}
\eta_{j,t,F1} &\sim \text{Normal}(0, \sigma_{j,YF1}) \\
\eta_{j,t,F2} &\sim \text{Normal}(0, \sigma_{j,YF2}),
\end{align}
where $\sigma_{j,YF1}$ and $\sigma_{j,YF2}$ determine year effect variability.

As in the growth sub-model (Appendix \ref{Colony growth}), Pearson correlation coefficients are computed from year effect estimates. For each sample of the posterior distribution, and for each species pair $(j,k)$, we calculate empirical correlations $\rho_{j,k,F1}$ and $\rho_{j,k,F2}$, respectively producing correlation matrices $\mathbf{R}_{F1}$ and $\mathbf{R}_{F2}$. These matrices capture how similarly different species respond to environmental fluctuations in terms of reproductive probability and egg production.

To generate random year effects for IPM simulations, we construct covariance matrices $\mathbf{\Sigma}_{F1}$ and $\mathbf{\Sigma}_{F2}$ from the correlation matrices and species-specific standard deviations:
\begin{align}
\Sigma_{j,k,F1} &= \sigma_{j,YF1} \, \sigma_{k,YF1} \, \rho_{j,k,F1} \\
\Sigma_{j,k,F2} &= \sigma_{j,YF2} \, \sigma_{k,YF2} \, \rho_{j,k,F2}.
\end{align}
Then, for each simulated year $t$, we sample the vectors of species-specific year effects $\boldsymbol{\eta}_{t,F1} = (\eta_{1,t,F1}, \ldots, \eta_{11,t,F1})$ and $\boldsymbol{\eta}_{t,F2} = (\eta_{1,t,F2}, \ldots, \eta_{11,t,F2})$ from multivariate normal distributions:
\begin{align}
\boldsymbol{\eta}_{t,F1} &\sim \text{MVN}(\mathbf{0}, \mathbf{\Sigma}_{F1}) \\
\boldsymbol{\eta}_{t,F2} &\sim \text{MVN}(\mathbf{0}, \mathbf{\Sigma}_{F2}).
\end{align}

To obtain the total number of eggs per colony, we must multiply the number of eggs per polyp, the number of polyps per unit area, and the colony area. Let $\text{polyps}_{i}$ denote the observed number of polyps per square centimeter for observation $i$ (n=35 colonies), and let $\theta_j$ denote the mean polyp density for species $j$. Using another Lizard Island dataset with 35 observations of polyp density, we fit the hierarchical model,

\begin{equation}
\theta_j \sim \text{Normal}(\mu_\theta, \sigma_\theta),
\end{equation}
\begin{equation}
\text{polyps}_i \sim \text{Normal}(\theta_{j[i]}, \sigma_{j[i]}),
\end{equation}
where $\mu_\theta$ and $\sigma_\theta$ are hyperparameters, and $\sigma_j$ represents the within-species standard deviation. 

One species, \textit{Acropora millepora}, lacked polyp density measurements. For this species, we used posterior samples from the fitted model to generate draws from:
\begin{equation}
\theta_{j^*} \sim \text{Normal}_{(0,\infty)}(\mu_{\text{Acr}}, \sigma_{\text{Acr}}),
\end{equation}
where $j^*$ indexes \textit{A. millepora}, and $\mu_{\text{Acr}}$ and $\sigma_{\text{Acr}}$ are the mean and standard deviation of $\theta_j$ across all other \textit{Acropora} species, for a single posterior sample. The $(0,\infty)$ subscript indicates a zero-truncated distribution.

For simulations of the IPM, we compute total egg production for each species by integrating across all size classes. For a colony of species $j$ and log planar area $x$, we first calculate the probability of being reproductive:
\begin{equation}
p_{j,t}(x) = \text{logit}^{-1}(\gamma_{j,0F} + \gamma_{j,1F} \cdot x + \eta_{j,t,F1}).
\end{equation}

For reproductive colonies, the expected number of eggs is calculated from the skew-normal parameters. When $X \sim \text{SN}(\xi, \omega, \alpha)$, the expectation of $\exp(X)$ (i.e., the mean egg count on the original non-logarithmic scale) can be derived from the moment-generating function of the skew-normal; given location $\xi_{j,t}(x) = \beta_{j,0F} + \beta_{j,1F} \cdot x + \eta_{j,t,F2}$, scale $\omega_j$, and shape $\alpha_j$, the expectation is
\begin{equation}
\mathbb{E}[\text{eggs} \mid \text{reproductive}=1] = \exp(\xi_{j,t}(x) + \omega_j^2/2) \cdot 2\Phi(\delta_j\omega_j),
\end{equation}
where $\delta_j = \alpha_j/\sqrt{1 + \alpha_j^2}$ and $\Phi$ is the standard normal cumulative distribution function.

Putting it all together, the average colony fecundity for species $j$ with log planar area $x$ is

\begin{equation}
F_{j,t}(x) = \overbrace{\exp(x)}^{\mathclap{\text{Colony area}}} \, \times \,  \underbrace{\theta_j}_{\mathclap{\text{Polyps per unit area}}} \, \times \,  \overbrace{p_{j,t}(x)}^{\mathclap{\text{Pr. reproductive}}} \, \times \,  \underbrace{\exp(\xi_{j,t}(x) + \omega_j^2/2) \, \times \,  2\Phi(\delta_j\omega_j)}_{\mathclap{\text{Conditional eggs per polyp}}}.
\end{equation}

\subsection{Recruitment} \label{Recruitment}

Recruitment rates depend on both relative egg production and available space. Let $n'_{j,t}(x)$ be the density of colonies (individuals per $\text{m}^2$) with log planar area $x$ (in $\text{cm}^2$), specifically after the survival and growth processes have occurred (hence the distinguishing ``prime'' notation). The total egg production per square meter is 
\begin{equation}
L_{j,t} = \int_{-\infty}^\infty n_{j,t}(x) F_{j,t}(x) dx.
\end{equation}
The proportion of total area occupied by species $j$ after the growth and survival processes is
\begin{equation}
N'_{j,t} = \int_{-\infty}^\infty n'_{j,t}(x) \exp(x) dx.
\end{equation}

With these two quantities, we can write the density of one-year recruits, denoted $R_{j,t+1}$ (individuals per m\textsuperscript{2}) as
\begin{equation} \label{eq:recruit1}
R_{j,t+1} = \overbrace{\frac{\beta_{j,R} L_{j,t}}{\sum_{k=1}^{11} L_{k,t}}}^{\mathclap{\text{\shortstack{Recruits per m$^2$ \\ on available substrate}}}} \times \underbrace{\left(1- \sum_{j=1}^{11} N'_{j,t}\right)}_{\mathclap{\text{Proportion open space}}},
\end{equation}
where $\beta_{j,R}$ determines the maximum recruitment rate. Density dependence arises through two mechanisms: larvae of all species compete for settlement space (the species-specific recruit density on available space is the relative abundance of eggs, scaled by $\beta_{j,R}$), and recruitment is limited to unoccupied space (represented by the factor $1- \sum_{j=1}^{11} N_{j,t}$).

The maximum recruit density parameters, $\beta_{j,R}$, cannot be directly estimated from data since recruitment was not observed. Instead, we apply a simulation-based procedure to select a range of $\beta_{j,R}$ that are consistent with biologically-realistic levels of coral cover.  For each species $j$ and each posterior sample of the other model parameters, we:
\begin{enumerate}
\item Select a sequence of $\beta_{j,R}$ values spanning from 10--1400 individuals m\textsuperscript{-2}.
\item For each candidate value, we simulate the species in isolation for 100 years, and then record mean coral cover over the subsequent $300$ years.
\item Randomly sample a target coral cover value from the interval $[0.1,0.5]$
\item Use linear interpolation to estimate the $\beta_{j,R}$ that produces this cover.
\end{enumerate}
This procedure represents our uncertainty in recruitment dynamics while maintaining realistic coral cover. The approximately linear relationship between $\beta_{j,R}$ and equilibrium cover justifies the interpolation approach. 

\begin{figure}[H]
\centering
\makebox[\textwidth]{\includegraphics[scale=0.9]{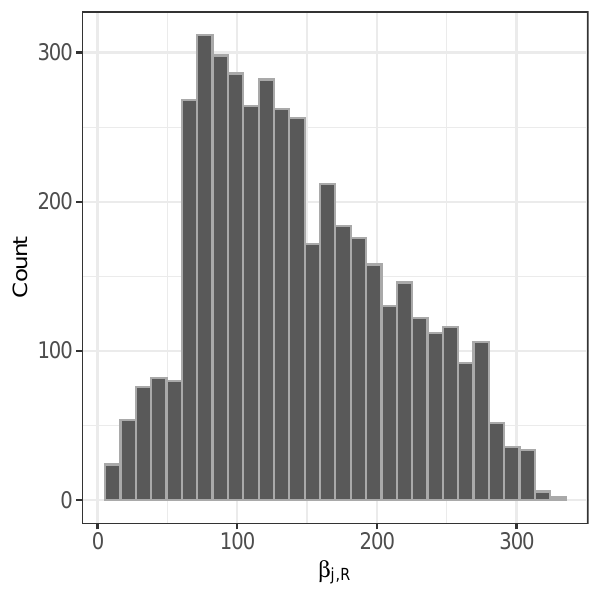}}
\caption{Distribution of maximum recruit density parameters, $\beta_{j,R}$. The distribution shows variation across species, simulation parameters, and target coral cover (between 0.1 and 0.5).}
\label{fig:max_recruit_dist}
\end{figure}

\begin{figure}[H]
\centering
\makebox[\textwidth]{\includegraphics[scale=0.9]{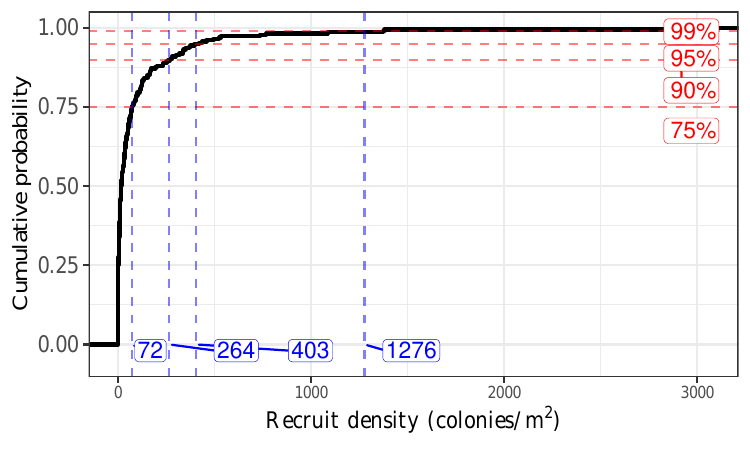}}
\caption{The empirical cumulative distribution function of one-year (or longer) coral recruit densities in Pacific Ocean sites. Data comes from \citet{edmunds2023coral}.}
\label{fig:recruit_ecdf}
\end{figure}

To validate our approach, we compared our estimated maximum recruit densities against empirical data from a recent meta-analysis of recruit densities \citep{edmunds2023coral}. The distribution of our estimated $\beta_{j,R}$ values is shown in Figure \ref{fig:max_recruit_dist}. The interquartile range of our estimates (77--199 individuals m\textsuperscript{-2}) corresponds to the 76th and 88th percentiles of observed one-year-or-longer recruit densities in the \cite{edmunds2023coral} meta-analysis, subset to Pacific ocean sites (Fig \ref{fig:recruit_ecdf}). While maximum recruit density cannot be directly observed in natural systems, the rough alignment between $\beta_{j,R}$ and large percentiles suggests that our simulation procedure produces biologically realistic values.

In IPM simulations, recruits are added based on a fixed size distribution. We assume recruit diameters follow a normal distribution with mean $\mu_D = 3.14$ cm and standard deviation $\sigma_D = 2.1$ cm, based on observed \textit{Acropora tenuis} recruit sizes from the northwest Philippines \citep{cruz2017enhanced}. To convert this to a distribution of log planar areas, we first convert diameter $d$ to area, assuming an idealized circular shape:

\begin{equation}
\text{area} = \pi(d/2)^2
\end{equation}

In the IPM simulations, colony size is necessarily discretized into bins of log planar area. For a given size class with boundaries $[x_L, x_U]$, the probability of a recruit falling in this class is:
\begin{equation} \label{eq:diam_CDF}
P(x_L \leq \log(\text{area}) \leq x_U) = \Phi\left(\frac{2\sqrt{\exp(x_U)/\pi} - \mu_D}{\sigma_D}\right) - \Phi\left(\frac{2\sqrt{\exp(x_L)/\pi} - \mu_D}{\sigma_D}\right),
\end{equation}
where $\Phi$ is the standard normal CDF. To find the actual number of recruits in each size category, we multiply the total number of recruits, $R_{j,t}$, by the size-specific probability mass above.

New coral recruits from a given year's spawning are not officially added to the colony size distribution until the following year. During their first year, recruits face different challenges than adults. Therefore, rather than applying standard model processes (growth, survival, and wave mortality), these young recruits' dynamics are captured through the density-dependent recruitment function (\eqref{eq:recruit2}) and the recruit size distribution.

\subsection{Complete model formulation} \label{Complete model formulation}

The complete integral projection model describes how size-structured coral populations change through time via survival, growth, and reproduction. Let $n_{j,t}(x)$ be the density of colonies of species $j$ with log planar area $x$ at time $t$. The population dynamics are governed by
\begin{equation}
n_{j,t+1}(y) = \underbrace{\int_{-\infty}^{\infty} G_{j,t}(y|x)S_{j,t}(x)n_{j,t}(x)dx}_{\text{survival and growth}} + \underbrace{R_{j,t}\phi(y)}_{\text{recruitment}}.
\end{equation}
The first term represents the survival and growth of existing colonies, while the second term captures the recruitment of new colonies; $\phi(y)$ is the size distribution of one-year recruits. Crucially, the new recruits were generated by adult colonies in the previous year (before the focal year's growth and survival processes), and are held in reserve until after the focal year's spawning event. Recall that the survival probability $S_{j,t}(x)$ combines both background mortality and disturbance-induced mortality.

The finite rate of increase, $\lambda_{j,t}$, is defined as the quotient of total cover between consecutive years, including recruits from each focal year:
\begin{equation}
\lambda_{j,t} = \frac{\int_{-\infty}^{\infty} \exp(x)[n_{j,t+1}(x) + R_{j,t+1}\phi_j(x)]dx}{\int_{-\infty}^{\infty} \exp(x)[n_{j,t}(x) + R_{j,t}\phi_j(x)]dx}
\end{equation}
The numerator represents total cover in year $t+1$, while the denominator represents total cover in year $t$. In each case, the term $\exp(x)$ converts from log planar area to actual area, and we sum across both existing colonies ($n_{j,t}(x)$) and recruits ($R_{j,t}\phi_j(x)$). The per capita growth rate $r_{j,t}$ is then be calculated as:

\begin{equation}
r_{j,t} = \log(\lambda_{j,t}).
\end{equation}

\subsection{Model-fitting details} \label{Model-fitting details}

We fit the baseline survival, growth, and fecundity models using \textit{Stan} \citep{carpenter2017stan}, a Bayesian statistical modeling platform, implemented through the \textit{rstan} package \citep{rstan} in R. To verify proper model convergence and sampling efficiency, we conducted standard diagnostic tests \citep[Ch. 6]{gelman2014bayesian}. All models met the following diagnostic criteria: the Gelman-Rubin statistic ($\hat{R}$) was below 1.1 for all parameters (indicating proper chain mixing), the effective sample size per iteration exceeded 0.001 (demonstrating efficient sampling), the energy Bayesian fraction of missing information (E-BFMI) was below 0.2 (suggesting appropriate model specification), and the proportion of divergent trajectories remained well below 1\% (indicating unbiased estimation).

For all parameters, we specified weakly informative prior distributions. To assess the relative influence of these priors, we calculated the posterior contraction for each parameter:
\begin{equation}
\text{post. contraction} = 1 - \frac{\mathbb{V}{\mathrm{post}}}{\mathbb{V}{\mathrm{prior}}}.
\end{equation}
This metric measures how much the posterior distribution has contracted relative to the prior distribution, with values closer to 1 indicating a stronger influence of the data relative to the prior \citep{schad2021toward}. Most parameters showed posterior contractions above 0.9, suggesting minimal prior influence. The main exceptions were the baseline-survival intercept parameters ($\beta_{j,0S}$), which showed stronger prior influence. 

The exceptions were some of the baseline-survival intercept parameters, . This higher prior influence putatively occurred because some species had few observations of colonies that were sufficiently small to produce a clear pattern of decrease in survival.

The complete diagnostic analysis, including diagnostic statistics and posterior contraction calculations for all models and parameters, is available in the supplementary files under {\fontfamily{qcr}\selectfont analysis/writeup/model\_diagnostics.Rmd}.

\subsection{IPM implementation} \label{IPM implementation}

To implement the integral projection model (IPM) numerically, we discretized the continuous state space of log planar area ($x$) into 40 equally spaced bins. All integrals were approximated using either the midpoint rule or the trapezoid rule. The domain of $x$ corresponds to a minimum area of 1 cm\textsuperscript{2}, and a maximum of 2.85 m\textsuperscript{2}. We enforced additional species-specific upper bounds on the maximum log colony area, calculated as the mean plus three standard deviations of observed log planar area in the Lizard Island growth dataset. These upper bounds were implemented because fecundity increases rapidly with area, meaning unrealistically large colonies could disproportionately influence results, though exploratory simulations with and without these constraints showed qualitatively similar outcomes—coexistence remained uncommon and \textit{G. pectinata} often excluded other species.

To address \textit{eviction} --- where probability mass is lost beyond state variable bounds --- we implemented a correction procedure using cumulative distribution functions to calculate and reassign probability mass that would fall outside the bounds to the respective minimum or maximum bins. Due to our liberal choice of size bounds, this approach did not result in artifactual accumulation of probability mass at the domain boundaries.

\section{Robustness analysis}
\label{Robustness analysis}

\subsection{Overview}

Here, we examine how our modeling choices affect scientific conclusions. Our primary finding --- the apparent weakness of the storage effect --- remains robust across multiple alternative models. In fact, these alternative models predict even weaker coexistence than our baseline model (Table \ref{tab:alternative_models}).

Each model variant is described in detail below. For most alternative models, we followed the methodology outlined in Appendix \ref{Recruitment} to recalculate the species-specific maximum recruit densities,$\beta_{j,R}$. However, for three scenarios —-- \textit{Equal max recruit density}, \textit{Low settlement probability}, and \textit{Variable settlement probability}  --- we fixed the maximum recruit density at $\beta_{j,R} = 250$ across all species.

% latex table generated in R 4.4.1 by xtable 1.8-4 package
% Fri Jan 17 08:49:57 2025
\begin{table}[H]
\centering
\caption{Probability of species coexistence across alternative models (rows) and community scenarios (columns). The \textit{Full community} scenario contains all 11 species, whereas the the \textit{Codominant} scenario contains only \textit{A. digitifera} and \textit{A. hyacinthus}.} 
\label{tab:alternative_models}
\begin{tabular}{l|cc|c}
\multicolumn{1}{l}{Model specification} & \rot{Pr. 2+ spp., Full community} & \rot{Pr. 3+ spp., Full community} & \rot{Pr. 2+ spp., Codominants} \\ 
 \hline
 Baseline simulation pars. & 0.38 & 0.12 & 0.16 \\ 
  No wave disturbance & 0.42 & 0.02 & 0.12 \\ 
  Const. radial growth model & 0.26 & 0.02 & 0.01 \\ 
  Generalized logistic survival & 0.40 & 0.09 & 0.13 \\ 
  Single-spp. cover, 0.4-0.5 & 0.32 & 0.08 & 0.15 \\ 
  Equal max recruit density & 0.30 & 0.06 & 0.10 \\ 
  Low settlement probability & 0.16 & 0.01 & 0.04 \\ 
  Variable settlement probability & 0.21 & 0.03 & 0.06 \\ 
  \end{tabular}
\end{table}

\subsection{No wave disturbance}
\label{No wave disturbance}

The wave-disturbance sub-model (Appendix \ref{Wave action survival}) presents two potential problems. First, since we estimate background mortality directly from demographic data and then add wave-disturbance mortality on top, we may be double-counting mortality if wave disturbances significantly impacted survival during the 6-year study period. Second, the wave-disturbance submodel may not be relevant for many reef zones. Large waves lose approximately 95\% of their height within the first 50 m of the reef crest \citep{madin2006scaling}, leaving the back reef and pre-crest reef slope much less susceptible to wave disturbances.

To address these concerns, we developed an alternative model that excludes wave-disturbance mortality entirely. To capture potential temporal variation in survival, we modified the background mortality model (Appendix \ref{Background mortality}) to include species-specific year effects. Specifically, we revised equation \eqref{eq:base_surv1} to
\begin{equation} 
\text{surv}_{i} \sim \text{Bernoulli}\left( \text{logit}^{-1}(\beta_{j[i],0S} + \beta_{j[i],1S} \cdot x_i + \eta_{j,t,S}\right),
\end{equation}
where $\eta_{j,t,S}$ is a species-specific year effect.

\subsection{Constant radial growth} \label{Constant radial growth}

The functional form of the original growth model (Appendix \ref{Colony growth}) was selected on the basis of graphical evidence, with the constraint that the model structure remains simple enough to avoid overfitting. For example, we use a student's $t$ error distribution on the basis of quantile-quantile plots, but enforce a linear relationship between colony growth and colony area  (after appropriate transformations). Here, we describe an alternative model that is more theoretically motivated: the \textit{constant radial growth model} \citep{madin2020partitioning}. This model assumes that the maximum colony growth is determined by a size-independent radial increment, with actual growth determined by the combined effects of the radial increment and partial mortality. The original growth model outperforms the constant radial growth model in leave-one-out cross-validation, but we examine the latter here as a robustness check.

For a colony with log planar area \( x_i \), we assume the colony is circular and therefore has the radius \( r_i = \sqrt{\exp(x_i)/\pi} \). Let \( \Delta r_i \) denote the observed radial increment (change in radius) between years \( t \) and \( t+1 \).

We model these increments using simple quantile regression for the chosen quantile \( \tau = 0.95 \) (for consistency with \citealp{madin2020partitioning}). Specifically, we assume that the quantile is time and size-independent:

\[
Q_{\Delta r_i}(\tau) = \alpha_{j[i]},
\]

where \( Q_{\Delta r_i}(\tau) \) represents the \( \tau \)-th quantile of the observed radial increment \( \Delta r_i \), and \( \alpha_{j[i]} \) represents the species-specific ``maximum'' increment. These parameters are estimated by minimizing the sum of the check function over all observations, implemented with the \textit{quantReg} R package \citep{quantreg}.

The radial growth increment determines the maximum area at time $t+1$, which can be written as 
\begin{equation}
a(x_i) = \pi\left(\sqrt{x_i/\pi} + \Delta r_i \right)^2,
\end{equation}

Let $y_i$ denote the observed log planar area after growth. The realized growth is simply $y_i = a(x_i)(1-p_i)$, where $p_i$ represents partial mortality. Solving for partial mortality gives
\begin{equation}
p_i = 1 - \frac{y_i}{a(x_i)}.
\end{equation}
For each observation in the data, we compute $p$ using this equation. We then model $logit(p_i)$ with a normal distribution.
\begin{equation}
\text{logit}(p_i) \sim \text{Normal}(\beta_{j[i],0G} + \beta_{j[i],1G} \cdot x_t, \sigma_{j[i],G}),
\end{equation}
where $\beta_{j,0G}$ and $\beta_{j,1G}$ are species-specific intercept and slope parameters, and $\sigma_{j[i],G}$ is the species-specific standard deviation. 

We used a two-step procedure to fit this model. In the first step, we estimated the $\alpha_{j}$ parameters using quantile regression. In the second step, we incorporated these parameters as fixed values into the Stan model to estimate partial mortality. We initially attempted to combine both aspects into a single Stan model (using the asymmetric laplace distribution to model the quantile), but this approach resulted in poor model-fitting diagnostics.

\begin{figure}[H]
\centering
\includegraphics[scale = 1]{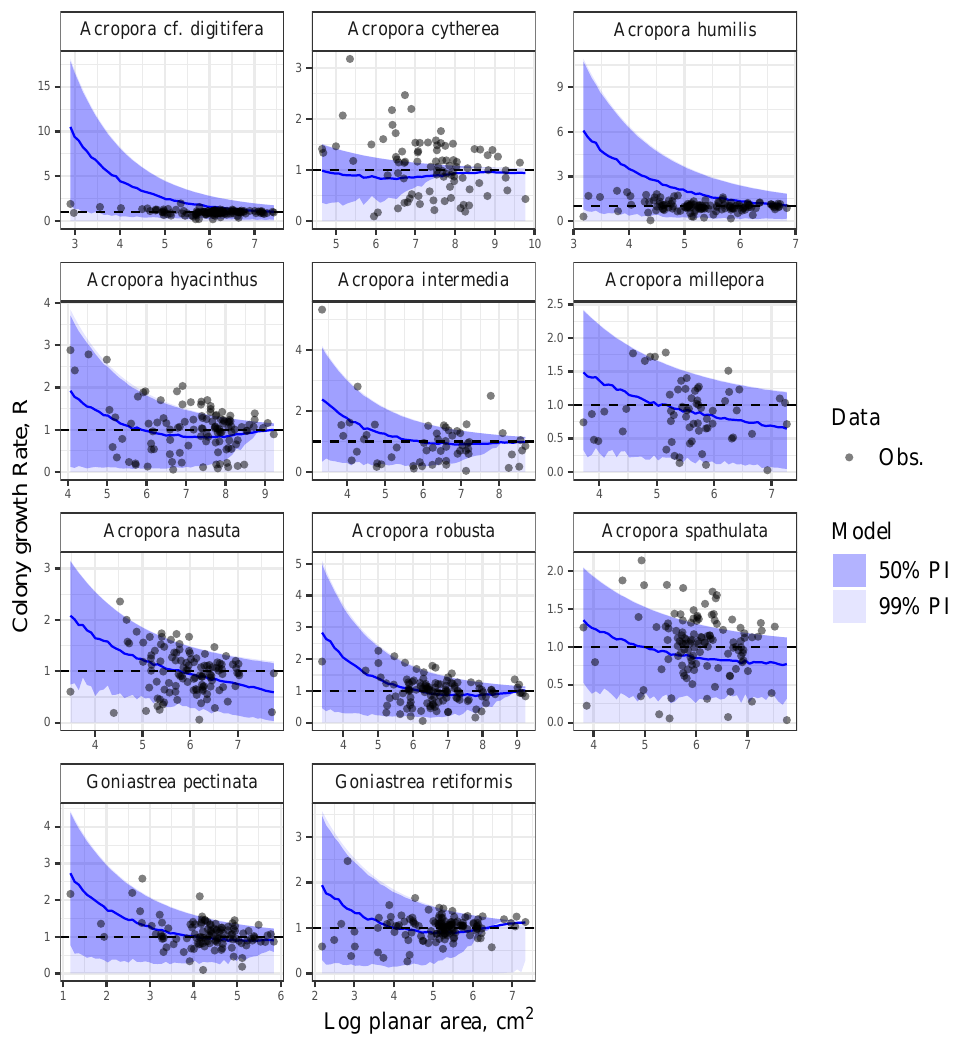}
\caption{Comparison between observed and predicted growth rates across colony size, with predictions generated by the \textit{constant radial growth} model. Each data point represents a single colony's growth in a particular year. The blue line represents the posterior mean, with the shaded regions showing predictive intervals (PI).}
\label{fig:growth_rad_fit}
\end{figure}

\subsection{Generalized logistic survival} \label{Generalized logistic survival}

The standard logistic model for baseline survival predicts near-zero survival rates for very small colonies. However, empirical studies show that 1-year coral recruits actually have survival rates between 5-20\% \citep{doropoulos2015linking, suzuki2018interspecific}. To avoid a biologically unrealistic scenario where the model might predict near-zero survival probabilities for some species' recruits, we developed a generalized logistic model that constrains survival probabilities between two parameters, $\alpha_{j[i],0S}$ and $\alpha_{j[i],1S}$. The statistic survival model (formerly \eqref{eq:base_surv1}) is thus

\begin{equation}
\text{surv}_{i} \sim \text{Bernoulli}\left(\alpha_{j[i],0S} + (\alpha_{j[i],1S} - \alpha_{j[i],0S}) \cdot \text{logit}^{-1}(\beta_{j[i],0S} + \beta_{j[i],1S} \cdot x_i)\right).
\end{equation}
For simulations of the IPM, the mortality fraction (formerly \eqref{eq:M}) is given by
\begin{equation} 
M_j(x) = 1 - \alpha_{j,0S} + (\alpha_{j,1S} - \alpha_{j,0S}) \cdot \text{logit}^{-1}(\beta_{j,0S} + \beta_{j,1S} \cdot x).
\end{equation}

The lower asymptote parameters ($\alpha_{j,0S}$) were given truncated normal priors on $[0,1]$ with mean 0.01 and standard deviation 0.1, while the upper asymptote parameters ($\alpha_{j,1S}$) were given truncated normal priors on $[0,1]$ with mean 0.99 and standard deviation 0.1. These priors reflect our biological understanding that very small colonies have low but non-zero survival probability, while very large colonies have high but not guaranteed background survival probability.
We chose not to use this more complex model in the main text because the asymptote parameters ($\alpha_{j[i],0S}$ and $\alpha_{j[i],1S}$) sometimes showed posterior contraction values well below 0.9, indicating that our choices of prior distributions are influencing posterior estimates.

\subsection{Diffuse correlations between species}
\label{Diffuse correlations between species}

Our main model handles interspecific correlations in growth and fecundity by fitting species-specific year effects independently, and then plugging these into Pearson's correlation estimator. This approach captures biologically realistic patterns where different species pairs show varying degrees of correlation in their environmental responses. However, since we don't formally model correlations or constrain them with priors, the estimated correlations likely show more variation than actually exists in nature. While a more formal model of pairwise correlations would be preferable (e.g., a Lewandowski-Kurowicka-Joe distribution) our dataset lacks the temporal replication that would be needed for this approach.

Here we explore an alternative \textit{diffuse correlation} model, where all species pairs share the same correlation. This only requires three additional parameters: one each for growth, reproductive probability, and egg production. Shrinkage with zero-centered beta priors was necessary to avoid divergent trajectories (evidence of biased estimation) in \textit{Stan}. The elements of the correlation matrixes, now in both the model-fitting process and the simulation model, are 
\begin{align}
\text{R}_{j,k,G} &=  \rho_{G} \\
\text{R}_{j,k,F1} &=  \rho_{F1}, \, \text{and}, \\
\text{R}_{j,k,F2} &=  \rho_{F2}, \\
\end{align}
for all $j \neq k$, and 1 otherwise.

\subsection{Single-spp. cover, 0.4-0.5}
\label{Single-spp. cover, 0.4-0.5}

In our main model, we randomly select each species' maximum recruitment density ($\beta_{j,R}$) to achieve equilibrium single-species coral cover between 0.1 and 0.5. This wide range could potentially create artificially large fitness differences between species. For a robustness check, we used a similar protocol but constrained species' equilibrium coral cover to a narrower range of 0.4 to 0.5.

\subsection{Equal maximum recruit density}
\label{Equal maximum recruit density}

Rather than selecting species-specific maximum recruitment densities, this alternative model sets a uniform value of $\beta_{j,R} = 250$ recruits per m\textsuperscript{2} for all species. This value was selected somewhat arbitrarily, but it tends to produce realistic coral cover (i.e., between 10\% and 60\%) for most species. Given the scarcity of data on species-specific recruitment probabilities, it is hard to say whether equal recruitment densities or variable recruitment densities (as presented in the main text) are more realistic.

\subsection{Low settlement probability}
\label{Low settlement probability}

The main model assumes high settler densities lead to density-dependent recruitment, without explicitly modeling the spawning-to-settlement transition. The alternative model here incorporates a fixed settlement probability ($p_L = 10^{-3}$, chosen based on literature values; Appendix \ref{Early life history transition probabilities}) and maximum recruitment density ($\beta_{j,R} = \beta_R = 250$). When total settler density falls below 250 individuals per m\textsuperscript{2}, density-dependent recruitment is bypassed and all settlers survive. The density of one-year recruits, previously given by \eqref{eq:recruit1}, becomes
\begin{equation} \label{eq:recruit2}
R_{j,t+1} = p_L \; L_{j,t}  \min\left(1, \frac{\beta_R}{\sum_{k=1}^{11} L_{k,t} p_L}\right) \left(1- \sum_{j=1}^{11} N'_{j,t}\right).
\end{equation}

To better understand how these dynamics differ from those in the baseline model, we can imagine a graph of recruit density as a function of settler density (for simplicity, consider a single species and hold the available area constant). The baseline model produces a horizontal line at the coordinate $y = \beta_{j,R}$. The low-settlement-probability model is a piecewise linear function, composed of the increasing line $y = x \cdot p_L$ when $x < \beta_R / (x \cdot p_L)$, then a horizontal line at $y = \beta_R$.

\subsection{Variable settlement probabilities}
\label{Variable settlement probabilities}

This alternative model builds on the \textit{Low settlement probability} model by introducing temporal variation in settlement success. Each year's settlement probability is drawn from a lognormal distribution, with $\log(p_{L,t}) \sim \text{Normal}(\mu = -3, \sigma = 2)$. Values are roughly based on available estimates (see Appendix \ref{Early life history transition probabilities}). Simulated values greater than 1 are clamped to $p_L = 1$.

\section{Early life history transition probabilities}
\label{Early life history transition probabilities}

This appendix serves two purposes. First, it highlights the substantial uncertainty in early life-history transition probabilities. This uncertainty underpins our main model's approach of bypassing explicit settlement probabilities in favor of assuming sufficient settler availability. Second, it establishes a reasonable spawning-to-settler transition probability for our robustness analysis (Appendix \ref{Low settlement probability}) where we relax this assumption. 

The early life history of corals involves two critical transitions: survival from spawning to competency, and successful settlement of competent larvae. For survival from spawning to competency (typically 3-7 days post-spawning), experimental estimates range from 50\% to 95\%. Settlement probability estimates show even greater variation, ranging from 5\% to 78\%. Combining these transition probabilities, the total probability of a spawned larva surviving to successful settlement ranges from 2.5\% (using 50\% survival × 5\% settlement) to 74.1\% (using 95\% survival × 78\% settlement) under various experimental conditions.

The early life history of corals involves two critical transitions: survival from spawning to competency, and successful settlement of competent larvae. For survival from spawning to competency (typically 3-7 days post-spawning), experimental estimates range from 50\% to 95\%. Settlement probability estimates show even greater variation, ranging from 5\% to 78\%. Combining these transition probabilities, the total probability of a spawned larva surviving to successful settlement ranges from 2.5\% (using 50\% survival × 5\% settlement) to 74.1\% (using 95\% survival × 78\% settlement) under various experimental conditions. 

Survival probability from spawning to competency shows strong consistency in the initial days post-spawning but varies considerably thereafter. \citet{graham2013rapid} demonstrated nearly 100\% survival for multiple species in the first 3 days, followed by species-specific decline rates. \citet{doropoulos2017density} reported approximately 95\% survival after 12 days in Acropora species. \citet{connolly2010estimating} provided detailed survival curves showing 20-60\% survival after 10 days, depending on species. All of these studies were conducted in laboratory conditions. \textit{In situ} survival probabilities rates are expected to be lower, since other (non-coral) planktonic marine larvae have estimated mortality rates of 10-25\% per day \citep{rumrill1990natural, lamare1999situ}.

Experimental settlement probability estimates cluster around different values depending on experimental conditions and measurement approaches. \citet{doropoulos2017density} observed approximately 45\% settlement probability for \textit{Acropora millepora} after 12 days using realistic larval densities of 250 individuals per liter. \citet{cruz2017enhanced} found approximately 15\% settlement probability for \textit{Acropora tenuis} in field conditions when accounting for available settlement area. \citet{edwards2015direct} found settlement probabilities for \textit{Acropora digitifera} from 5--78\% depending on the method of calculation (see \hyperref[sec:settlement-notes]{Settlement probability notes} below). \citet{suzuki2012optimal} found settlement rates between 9\% and 15\% across different larval densities of two Acropora species.

Compared to experiments, coral models predict substantially lower spawning-to-settlement probabilities. In their integral projection model (IPM), \citet{alvarez2023disturbance} chose spawning-to-recruit (1-year) probabilities of 0.00000257 and 0.00001 for two \textit{Acropora} species, because these values successfully reproduced realistic coral cover levels (65\%). When accounting for one-year recruitment probabilities of 5-20\% \citep{suzuki2018interspecific}, the implied spawning-to-settlement probabilities range from 1.3$\times$10\textsuperscript{-5} to 2.0$\times$10\textsuperscript{-4}. Other IPM studies have explored broader ranges of spawning-to-settlement probabilities, with \citet{bairos2023demographic} finding that values from $10^{-4}$ to $10^{-1}$ produced realistic coral cover in models with density-dependent recruitment. Similarly, \citet{mcwilliam2023net} identified $10^{-4}$ to $10^{-1}$ as the range of plausible spawning-to-recruitment probabilities.

The disparity between experimental and theoretical estimates can be largely attributed to larval retention dynamics. Experiments necessarily constrain larvae near settlement surfaces, whereas in natural systems, larvae are subject to dispersal by oceanic currents. Hydrodynamic models indicate that larval retention rates on natal reefs vary dramatically, ranging from 0 to 60\% depending on reef morphology, with higher retention in lagoons and archipelagos compared to isolated reefs \citep{andutta2012sticky}. The precision of retention estimates is further complicated by model uncertainty \citep{choukroun2024larval} and substantial temporal variability. For instance,  demonstrated that retention rates can fluctuate by up to 40\% within just the first 48 hours post-spawning, driven by variations in tidal and wave forcing \citep{grimaldi2022hydrodynamic}. This complexity and uncertainty makes it difficult to develop an accurate sub-model for settlement.

\subsection{Settlement probability calculations}\label{sec:settlement-notes}

This section details our approach to calculating some settlement probabilities used in the previous section. While some papers don't explicitly state these probabilities, they do provide the necessary data for their calculation.

In \citet{cruz2017enhanced}, researchers placed $\sim$400,000 \textit{Acropora tenuis} larvae in each of four 6m $\times$ 4m larval enhancement plots (24m\textsuperscript{2} each plot), with ten 10cm $\times$ 10cm settlement tiles per plot. Initial settlement counts found 1,021 spat total across all tiles in all plots. To calculate the area-adjusted settlement probability, we first determine that the 40 total tiles (0.01m\textsuperscript{2} each) covered 0.4m\textsuperscript{2} out of the total 96m\textsuperscript{2} plot area (4 plots $\times$ 24m\textsuperscript{2}), or 0.417\% of available space. Assuming uniform larval distribution, of the 1.6 million total larvae released (400,000 $\times$ 4 plots), approximately 6,672 larvae (1.6 million $\times$ 0.00417) would have encountered the tile surfaces. With 1,021 settled spat observed, we obtain a settlement probability of 15.3\% (1,021/6,672). This calculation assumes uniform larval distribution across all surfaces, equal settlement preference between tiles and natural substrate, and equal accessibility of all tile surfaces.

In \citet{edwards2015direct}, researchers conducted a large-scale larval seeding experiment where they released $\sim$1,040,000 \textit{Acropora digitifera} larvae onto seven artificial reef structures (``pallet balls''), each with 2.94m\textsuperscript{2} of surface area. Settlement density was measured by counting spat, finding an average of 257.2 spat per 0.1m\textsuperscript{2}. Two methods were used to calculate settlement probability: The first method simply divided the total number of settled spat (257.2 spat/0.1m\textsuperscript{2} $\times$ 205.8 [total area in 0.1m\textsuperscript{2} units] = 52,931 spat) by the total larvae released (1,040,000), producing 5.09\%. The second method accounted for the study area footprint -- since the pallet balls only occupied 6.51\% of the total study area (7.91m\textsuperscript{2} out of 121.44m\textsuperscript{2}), only about 67,704 larvae (6.51\% of 1,040,000) would be expected to encounter the settlement surfaces. Using this new denominator gives a higher settlement probability of 78.18\% (52,931/67,704). The difference between these estimates (5.09--78.18\%) demonstrates how settlement probability calculations are highly sensitive to assumptions about larval distribution and available settlement area.

\section{Modern coexistence theory} \label{Modern coexistence theory}

\subsection{Overview} 

Modern coexistence theory (MCT) provides a framework for measuring the relative importance of different explanations for coexistence. The fundamental approach can be summarized as \textit{decompose and compare}. We \textit{decompose} each species' long-term per capita growth rate into additive components. Because stable coexistence occurs when there is a rare-species advantage, we \textit{compare} the components of rare species (invaders) and common species (residents). The invader-resident comparison is essential because a mechanism can promote coexistence by helping rare species or hurting common species.

Classic MCT uses a Taylor series to decompose per capita growth rates. Each species $j$ has a per capita growth rate function $r_j(E_j,C_j)$, which depends on species-specific environmental responses $E_j$ and competition $C_j$. One then defines equilibrium parameters that satisfy $r_j(E_j^*,C_j^*) = 0$, and performs a Taylor series decomposition about the equilibrium \citep{barabas2018chesson}. An alternative formulation first uses a change of coordinates $\mathcal{E}_j = r_j(E_j,C_j^*)$ and $\mathcal{C}_j = -r_j(E_j^*,C_j)$ \citep{chesson1994multispecies}. The additive Taylor series terms can be grouped in different ways, and certain assumptions are needed to ensure that third-order and higher terms are of negligible magnitude (\citealp{chesson1994multispecies}, Appendix II).

Classic MCT has led to many theoretical insights (e.g., \citealp{chesson1997roles, stump2015distance, schreiber2021positively}), but it faces several challenges as an empirical tool for ``measuring coexistence'' in the real world. First, the conventional assumption of small environmental noise often doesn't hold in real ecosystems. When environmental noise is substantial, the higher-order terms that MCT typically ignores may actually play a crucial role in determining species' growth rates. Second, classic MCT typically finds that environmental and competitive fluctuations have comparable effects, which stems from its assumptions of small environmental noise and a shared competition parameter. When these assumptions don't hold, traditional mechanisms like resource partitioning may dominate over storage effects \citep{stump2023reexamining}. Third, complicated community models lead to laborious mathematical calculations --- a challenge that becomes particularly evident in structured population models. In structured populations, even first-order terms in the Taylor-series decomposition require complex expressions involving eigenvalue sensitivities (e.g., $\frac{\partial r}{\partial E} = \frac{1}{\lambda} \sum_{i,j} v_i w_j \frac{\partial a_{ij}}{\partial E}$); this complexity obscures biological interpretation. Fourth, finding the equilibrium population density function $n^*(z)$ for the Taylor expansion becomes problematic with structured populations, as we must solve for stable size distributions through eigenvalue problems rather than simple scalar equilibria.

To overcome these limitations, we use \textit{simulation-based MCT} \citep{ellner2016data, Ellner2019, johnson2023coexistence}, which does away with analytical expressions of Taylor series terms, and instead directly calculates coexistence mechanisms with simulated data and a population growth function.

\subsection{Growth rate partitioning} \label{Growth rate partitioning}

For a per capita growth rate function $r_j(E_j,C_j)$, its long-term average is decomposed as 

\begin{equation} \label{eq:coarseMCT2}
\overline{r_j(E_j,C_j)} = \varepsilon_j^0 + \varepsilon_j^E + \varepsilon_j^C + \varepsilon_j^{EC}
\end{equation}

where:

\begin{itemize} 
\item $\varepsilon^0 = r(\bar{E},\bar{C})$ is the baseline growth rate at mean environment and competition levels,
\item $\varepsilon^E = \overline{r(E,\bar{C})} - \varepsilon^0$ captures the main effect of environmental variation,
\item $\varepsilon^C = \overline{r(\bar{E},C)} - \varepsilon^0$ captures the main effect of competition variation, and 
\item $\varepsilon^{EC} = \overline{r(E,C)} - [\varepsilon^0 + \varepsilon^E + \varepsilon^C]$ represents the interaction between environment and competition.
\end{itemize}

The interaction term $\varepsilon^{EC}$ can be further decomposed into independent and covarying components:
\begin{equation} \label{eq:epsEC}
\varepsilon^{EC} = \varepsilon^{(E\#C)} + \varepsilon^{(EC)}
\end{equation}
Here, $\varepsilon^{(EC)}$ captures the effect of environment-competition covariance, and $\varepsilon^{(E\#C)}$ measures the independent effects of variation in $E$ and $C$. To isolate these independent effects in $\varepsilon^{(E\#C)}$, we use a randomization approach. First, we collect $E_j$ and $C_j$ from simulations. Then we randomly shuffle the $E_j$ values to eliminate any correlation with competition. Using these shuffled environmental values and the original competition values, we recalculate growth rates and average them over time. Using the original $C$ values ensures that the storage effect term only reflects $E$-$C$ covariation (and higher-order interactions), not the marginal effects of variation in $C$.

The $\varepsilon^{(EC)}$ term will become the storage effect after the invader--resident comparison, though the correspondence between this simulation-based storage effect and the classic-MCT storage effect is not exact (\citealp{Ellner2019}, Appendix 3). The term $\varepsilon^{(E\#C)}$ captures higher-order interaction terms that are negligibly small in Classic MCT (though not necessarily here). 

To assess how each component contributes to coexistence, we compute invader-resident comparisons. For example, the comparison of the mean effect for invader $i$ is
\begin{equation} \label{eq:inv_res_comparison1}
\Delta_i^0 = \varepsilon_{i|i}^0 - \frac{1}{S-1}\sum_{r\neq i} \varepsilon_{r|i}^0,
\end{equation}
where $r$ indexes resident species, $S$ is the number of species (therefore, $S-1$ is the number of residents), and $\varepsilon_{j|i}^0$ represents the baseline growth rate for species $j$ when species $i$ is invading. This comparison method can be applied to any of the $\varepsilon$ terms in equation \ref{eq:coarseMCT2} or \ref{eq:epsEC}. Because each resident is weighted equally (dividing by $S-1$) the differencing scheme in \eqref{eq:inv_res_comparison1} is called the simple comparison \citep{johnson2022methods}.

\begin{figure}[H]
\centering
\includegraphics[scale = 1]{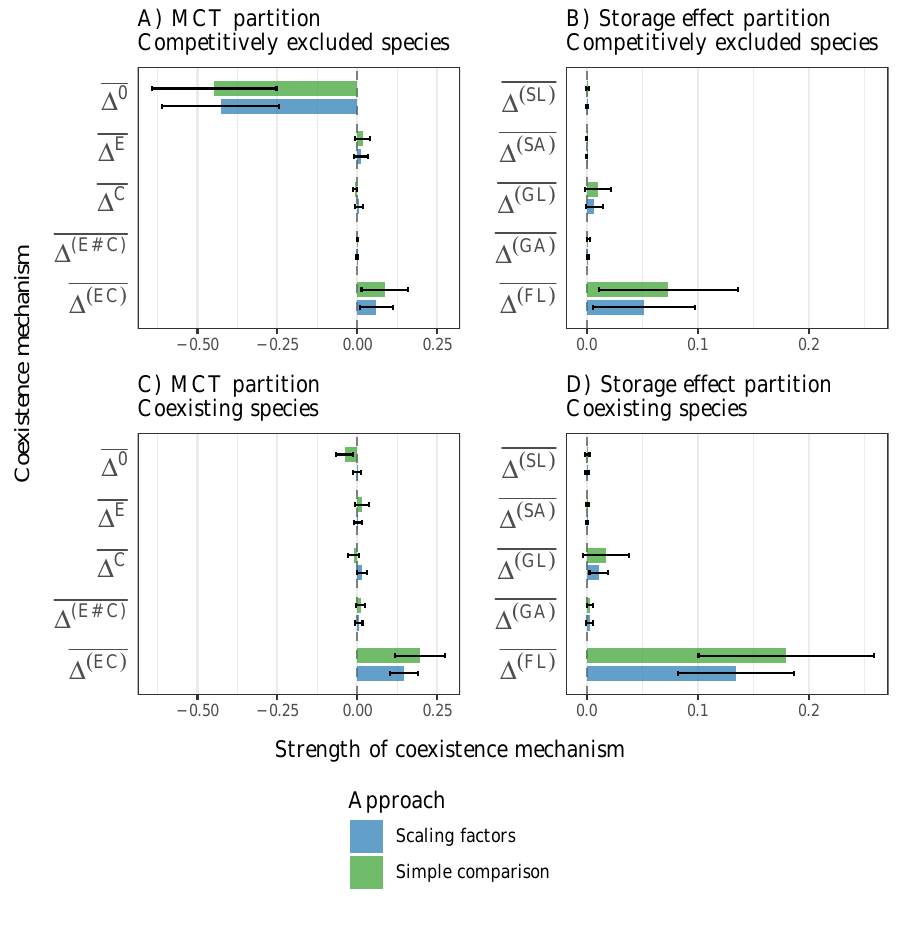}
\caption{The strength of coexistence mechanisms remains consistent whether measured by the \textit{simple comparison} or \textit{scaling factors} method. Panels show mechanism strength across competitively excluded species (top row), coexisting species (bottom row), the coarse-grained MCT partition (left column), and the fine-grained MCT partition (right column). Bars and error bars respectively show the Mean and SE across both community modules (e.g., all species, just tabular species) of the posterior means.}
\label{fig:scaling_factor_bars}
\end{figure}

Classic MCT computes the invader--resident comparison slightly differently, using \textit{scaling factors} to reweight the resident growth rate components. We computed these scaled mechanisms using the Simulation-regression method described in \citet{ellner2016quantify}, Appendix B. Specifically, we defined $\mathcal{C}_j = -r_j(\overline{E_j}, C_j)$ and performed a multiple regression with the invader's $\mathcal{C}_i$ as the response variable and the residents' $\mathcal{C}_r$ as predictors. The resulting regression slopes serve as the residents' scaling factors. Since both the scaling factors and the simple comparison methods produced similar results (Fig. \ref{fig:scaling_factor_bars}), we use the simple comparison method throughout the remainder of our analysis.

It is important to note that in general, the $\Delta_i^{0}$ term cannot be strictly equated with fitness differences for several reasons. First, it incorporates both equalizing and stabilizing components that arise from species interactions in the absence of environmental fluctuations (such as resource partitioning). Second, the term "fitness differences" has already been used to describe other quantities in modern coexistence theory \citep{song2019consequences}. Third, the term reflects indirect effects of fluctuations --- if competition is a nonlinear function of the environment (i.e., $C = h(E)$), then $\overline{C(E)} \neq C(\overline{E})$ via Jensen's inequality. While the second and third points do apply to our coral model (suggesting an implicit asterisk should accompany our use of "fitness differences"), these concerns are not major concerns. Overlap in terminology and inexact correspondence between conceptual objects (here, fitness differences) and their measurements (here, $\Delta_i^{0}$) are common in ecological theory, and the term remains useful as a conceptual shorthand.

Finally, we define the community-average coexistence mechanisms by averaging the invader--resident differences across all possible invader configurations for a given resident community. For example, the community average of the main effect of $E$-variation is
\begin{equation} \label{eq:com_average1}
\overline{\Delta^E} = \frac{1}{n_{\text{inv}}}\sum_{i=1}^{n_{\text{inv}}} \Delta_i^E
\end{equation}
where $n_{\text{inv}}$ is the total number of possible invader configurations. For example, in our 11-species coral community, when only a single species persists, $n_{\text{inv}}$ equals 10 (the other 10 species invading individually). For a two-species resident community, $n_{\text{inv}} =  11$: nine species that can invade from outside the resident community, plus each of the two resident species acting as invaders while the other maintains its equilibrium density. These community-average mechanisms capture the overall strength of coexistence mechanisms across all possible invader-resident combinations for a given resident community structure. 

\subsection{Coarse-grained MCT partition}
\label{Coarse-grained MCT partition}

To implement simulation-based MCT, we first define environmental and competition parameters. Competition is measured as the logarithm of total egg production relative to available space:
\begin{equation}
C_{t} = \log\left(\frac{\sum_{k=1}^{11} L_{k,t}}{1-\sum_{k=1}^{11} N'_{k,t}}\right),
\end{equation}
where $L_{k,t}$ is the total egg production of species $k$ and $N'_{k,t}$ is the proportion of area occupied by species $k$ after growth and survival processes. Note that all species share a common competition parameter.

The environmental parameter for species $j$ is defined as a set containing the dislodgement mechanical threshold and all species-specific year effects:
\begin{equation} \label{eq:E_set}
E_{j,t} = \Big\{ \text{DMT}_t, \; \eta_{j,t,G}, \; \eta_{j,t,F1}, \; \eta_{j,t,F2} \Big\},
\end{equation}
where $\text{DMT}_t$ captures wave disturbance intensity, $\eta_{j,t,G}$ represents the year effect on growth, and $\eta_{j,t,F1}$ and $\eta_{j,t,F2}$ represent year effects on reproductive probability and egg production respectively.

Setting the environmental parameter to its temporal mean, denoted $\overline{E_j}$, is equivalent to
\begin{equation}
E_{j,t} = \overline{E_j} \iff \Big\{ \text{DMT}_t = \overline{\text{DMT}}, \; \eta_{j,t,G} = 0, \; \eta_{j,t,F1} = 0, \; \eta_{j,t,F2} = 0 \Big\}.
\end{equation}

When shuffling $E_{j,t}$ (to calculate $\varepsilon_j^{E\#C}$) we treat the set of environmental variables (i.e., the RHS of \eqref{eq:E_set}) as a single unit, preserving any covariation between demographic rates in order to measure the overall ``coarse-grained'' storage effect. Specifically, when shuffling environmental conditions to compute $\varepsilon_j^{E\#C}$, all elements of $E_{j,t}$ from a given time point are kept together, preserving the temporal correlation structure between environmental variables while breaking their correlation with competition. 

The per capita growth rate $r_{j,t}$ is defined as the logarithm of the finite rate of increase:
\begin{equation} \label{eq:r1}
r_{j,t} = \log\left(\frac{\int_{-\infty}^{\infty} \exp(x)[n_{j,t+1}(x) + R_{j,t+1}\phi_j(x)]dx}{\int_{-\infty}^{\infty} \exp(x)[n_{j,t}(x) + R_{j,t}\phi_j(x)]dx}\right),
\end{equation}
where $n_{j,t}(x)$ represents the density of colonies with log planar area $x$, $R_{j,t}$ is the recruitment density, and $\phi_j(x)$ is the size distribution of recruits.

To show how our coral model aligns with the formalism of MCT, we can make explicit the dependence on the environment and competition:

\begin{equation} \label{eq:r2}
r_j(E_{j,t}, C_{t}, n_{j,t}, R_{j,t}) = \log\left(\frac{\int_{-\infty}^{\infty} \exp(x)\left[n_{j,t+1}(x|n_{j,t},E_{j,t}) + R_{j,t+1}(E_{j,t},C_{t})\phi_j(x)\right]dx}{\int_{-\infty}^{\infty} \exp(x)\left[n_{j,t}(x) + R_{j,t}\phi_j(x)\right]dx}\right),
\end{equation}
where $n_{j,t+1}(x|n_{j,t},E_{j,t})$ indicates that next year's size distribution depends on both the current size distribution and environmental conditions, while $R_{j,t+1}(E_{j,t},C_{t})$ shows that next year's recruitment depends on current environmental conditions and competition. 

Our approach to calculating per capita growth rates may seem unconventional, since we combine coral cover measurements from different points in time. For example, in the numerator of equation \eqref{eq:r1}, we add together $n_{j,t+1}$ (adult colony cover at the start of the next time step) and $R_{j,t+1}$ (recruitment at the end of the next time step). The conceptual argument for this is that both of these measurements share an important characteristic: both are determined by processes from the previous time step and are unaffected by processes in the subsequent time step. The new recruits, though not physically added to the population until later, are immune from fluctuating environmental conditions.

Functionally, this formulation of per capita growth rates helps us analyze the storage effect. It may be more intuitive to think of a simulation where events are reordered (spawning, then survival, growth, and recruitment), such that growth rate calculations now only involve adult cover (because one-year recruits were just added to adult populations). However, in this formulation, environmental fluctuations affecting survival and growth affect competition in subsequent time steps, causing the storage effect to be absorbed into MCT's relative nonlinearity term (i.e., $\Delta^C$ in the simulation-based MCT). This shift would split storage effects across multiple terms --- survival and growth effects would appear in $\Delta^C$, while fecundity fluctuation effects would remain in $\Delta^{(EC)}$. Such a division makes it challenging to isolate and measure the storage effect independently.

\subsection{Fine-grained MCT partition}
\label{Fine-grained MCT partition}

We can further decompose the storage effect term ($\varepsilon_j^{(EC)}$) by recognizing that both environmental and competition parameters consist of multiple components. The environmental parameter for species $j$ can be written as a set containing survival, growth, and fecundity components:
\begin{equation}
E_j = \Big\{S_j, G_j, F_j\Big\},
\end{equation}
where:
\begin{align}
S_j &= \text{DMT}_t && \text{(wave disturbance intensity)} \\
G_j &= \eta_{j,t,G} && \text{(growth year effect)} \\
F_j &= \Big\{ \eta_{j,t,F1}, \eta_{j,t,F2} \Big\} && \text{(fecundity year effects)}
\end{align}
Similarly, the competition parameter can be decomposed into two components:
\begin{equation}
C = \Big\{L, A\Big\},
\end{equation}
where:
\begin{align}
L &= \log\left(\sum_{k=1}^{11} L_{k,t}\right) && \text{(total larval production)} \\
A &= \log\left(\sum_{k=1}^{11} N'_{k,t}\right) && \text{(total occupied area)}
\end{align}

With this more detailed parameterization, we can write the per capita growth rate as a function of five variables:
\begin{equation}
r_j(S_j, G_j, F_j, L, A)
\end{equation}
This formulation allows us to decompose the growth rate into zeroth-order (mean), first-order (main effects), second-order (pairwise interactions), and higher-order terms:

\begin{align}
\overline{r_j(S_j, G_j, F_j, L, A)} &= \underbrace{\varepsilon_j^0}_{\text{mean}} + \nonumber \\
& \underbrace{\varepsilon_j^S + \varepsilon_j^G + \varepsilon_j^F + \varepsilon_j^L + \varepsilon_j^A}_{\text{main effects}} + \nonumber \\
& \underbrace{\varepsilon_j^{SL} + \varepsilon_j^{SA} + \varepsilon_j^{GL} + \varepsilon_j^{GA} + \varepsilon_j^{FL} + \varepsilon_j^{FA}}_{\text{pairwise interactions}} + \nonumber \\
& \underbrace{\varepsilon_j^{\text{``HOT''}}}_{\text{higher-order terms}}
\end{align}

To illustrate how these terms are calculated, consider the first-order effect of survival ($\varepsilon_j^S$):
\begin{equation}
\varepsilon_j^S = \overline{r_j(S_j, \overline{G_j}, \overline{F_j}, \overline{L}, \overline{A})} - \epsilon_j^0
\end{equation}
Similarly, a second-order interaction term like $\varepsilon_j^{SL}$ (the interaction between survival and larval production) is calculated as:
\begin{equation}
\varepsilon_j^{SL} = \overline{r_j(S_j, \overline{G_j}, \overline{F_j}, L, \overline{A})} - [\varepsilon_j^0 + \varepsilon_j^S + \varepsilon_j^L]
\end{equation}
The higher-order terms ($\varepsilon_j^{\text{``HOT''}}$) capture all remaining interactions not accounted for by first and second-order terms:
\begin{equation}
\varepsilon_j^{\text{``HOT''}} = r_j(S_j, G_j, F_j, L, A) - [\text{all zeroth, first, and second-order terms}]
\end{equation}

By shuffling individual environmental parameters while keeping others fixed, we can calculate ``sub-storage effects'' such as $\varepsilon_j^{(FL)}$, which captures the contribution of fecundity-larvae covariance to the overall storage effect. Two technical notes: First, the sub-storage effects do not sum exactly to the storage effect from the coarse-grained partition ($\varepsilon_j^{(EC)}$). This is because the coarse-grained storage effect captures situations where multiple environmental or competition variables covary simultaneously. Second, the fecundity-area interaction term ($\varepsilon_j^{FA}$) has an expected value of zero because fecundity during spawning cannot affect adult coral cover within the same time step.

The five biologically relevant storage effects in our system are $\varepsilon_j^{(SL)}$, $\varepsilon_j^{(SA)}$, $\varepsilon_j^{(GL)}$, $\varepsilon_j^{(GA)}$, and $\varepsilon_j^{(FL)}$, representing the covariance between survival and larvae, survival and area, growth and larvae, growth and area, and fecundity and larvae, respectively (Fig. \ref{fig:conceptual_fig}). To calculate these sub-storage effects, we use a shuffling approach similar to that described for the coarse-grained partition. For example, to calculate $\varepsilon_j^{(FL)}$, we:

\begin{enumerate}
\item Collect time series of all environmental and competition variables from simulation
\item Shuffle the temporal ordering of the sets of fecundity year effects, $F_{j,t} = \Big\{ \eta_{j,t,F1}, \eta_{j,t,F2} \Big\}$.
\item Calculate per capita growth rates using shuffled $F_j$ and original $L$, with all other variables (i.e., $S, G, A$) held at their temporal averages. That is, we do not not use the new per capita fecundity $F_j$ (after shuffling) to recompute the total larval density $L$.

\item Take the temporal average of per capita growth rates, completing the calculation of $\varepsilon_j^{(F\#L)}$
\item Compute the difference, i.e., $\varepsilon_j^{(FL)} = \varepsilon_j^{FL} - \varepsilon_j^{(F\#L)}$
\end{enumerate}

As with the coarse-grained partition, we complete the fine-grained partition by computing invader-resident comparisons and averaging over all possible invader configurations (see \eqref{eq:inv_res_comparison1} \& \eqref{eq:com_average1}).

\subsection{Invasion analysis algorithm}
\label{Invasion analysis algorithm}

Starting with all species present, we allow the community to attain its stationary distribution. We then systematically analyze invasions by removing each species in turn, allowing the resulting subcommunity to equilibrate, and then measuring the invasion growth rate of the removed species. These growth rates are then decomposed using MCT.

\begin{algorithm}[H]
\footnotesize  
\caption{Modern Coexistence Theory Analysis}
\label{alg:mct}

\begin{enumerate}
    \item \textbf{Parameter Sampling}
    \begin{enumerate}
        \item Draw $200$ sets of model parameters from the joint posterior distribution
        \item For each parameter set, execute steps 2--4
    \end{enumerate}

    \item \textbf{Initial Community Assembly}
    \begin{enumerate}
        \item Initialize the model with all 11 species by adding recruits. Total initial cover is 50\%, split evenly between species.
        \item Run simulation for $10$ years with small immigration rate (2 individuals m\textsuperscript{-2} per year), to potentially help setup a stable size structure.
        \item Run simulation for additional $1000$ steps without immigration
        \item Identify coexisting species at the end of this period. Species are considered extirpated if their cover ever drops below $10^{-5}$.
    \end{enumerate}

    \item \textbf{Invasion Analysis Subcommunities}
    \begin{enumerate}
        \item If multiple species coexist, include all 11 species as invaders in the  invasion analysis (step 4).
        \item If only one species coexists, only treat the 10 non-coexisting species as invaders in step 4; i.e, skip the ecologically-irrelevant analysis of a single species invading an empty community.
    \end{enumerate}

    \item \textbf{Invasion Analysis} \\
    For each potential invader species $i$:
    \begin{enumerate}
        \item \textbf{Resident Community Preparation}
        \begin{enumerate}
            \item Remove species $i$ if currently present
            \item Run simulation for $500$ years to allow resident community to equilibrate
        \end{enumerate}
        
        \item \textbf{Invasion Trial Setup}
        \begin{enumerate}
            \item Introduce invader at low density (initial cover $10^{-15}$, all in the recruit stage)
            \item Run simulation for $20$ steps to establish stable size structure
        \end{enumerate}
        
        \item \textbf{Data Collection Phase}
        \begin{enumerate}
            \item Run simulation while collecting data for MCT partitions
            \item Monitor invader population:
            \begin{itemize}
                \item If invader goes extinct (numerical underflow): repeat ``invasion trial setup''.
                \item If invader becomes too common (cover $> 10^{-5}$): remove and reintroduce at low density
            \end{itemize}
            \item Continue until $500$ years of data are collected
        \end{enumerate}
        
        \item \textbf{MCT Analysis}
        \begin{enumerate}
            \item Calculate coarse-grained partition terms following Section \ref{Coarse-grained MCT partition}
            \item Calculate fine-grained partition terms following Section \ref{Fine-grained MCT partition}
        \end{enumerate}
    \end{enumerate}

    \item \textbf{Data Processing and Analysis}
    \begin{enumerate}
        \item Compute invader-resident comparisons for each mechanism (e.g., \eqref{eq:inv_res_comparison1})
        \item Calculate community-average mechanisms across all invader configurations (e.g., \eqref{eq:com_average1}).
        \item Aggregate results across all posterior samples: compute means and standard errors.
        \item For analyses that compare coexistence mechanisms in non-coexistence scenarios vs coexistence scenarios, repeat steps (b) and (c) above separately for two groups:  invaders that can't coexist, invaders that can coexist.
    \end{enumerate}
\end{enumerate}
\end{algorithm}

\subsection{Connecting growth rates to coexistence}
\label{Connecting growth rates to coexistence}

The \textit{mutual invasibility criterion} is the traditional approach for relating growth rates to coexistence, though it has important limitations. The \textit{mutual invasibility criterion} \citep{macarthur1967limiting, turelli1978does} is a heuristic which states that species coexist if each species can invade while all $S-1$ remaining species are at their typical densities (i.e., a positive stationary distribution). While this criterion proves reliable for two-species communities (given minimal conditions like the absence of Allee effects; \citealp{chesson1989invasibility}) and for scenarios with similar interaction strengths (\textit{diffuse-competition}; \citealp{chesson2000mechanisms}), it encounters significant limitations in species-rich communities. These limitations arise because not all $S-1$ subcommunities may exist, making invasion analysis impossible. Additionally, the criterion fails to provide guidance when only a subset of species ultimately coexist.

A new framework \citep{spaak2023building} builds on permanence theory to more rigorously analyze coexistence. This framework examines invasion growth rates across all possible subcommunities and creates an invasion graph. In cases where this invasion graph is acyclic (lacking cycles), coexistence is guaranteed when two conditions are met: each subcommunity must be susceptible to invasion by at least one absent species, and during transitions between subcommunities, species that join the community must be capable of invasion while those that leave have negative invasion growth rates.

While our invasion analysis (Appendix \ref{Invasion analysis algorithm}) appears to differ from \citepos{spaak2023building} framework, it is actually a special case of their method that is valid for our specific system.  For the vast majority of posterior samples in our coral models, only 1--2 species can persist/coexist in our model. This has three important implications: all $S-1$ subcommunities exist and are stable, community assembly cycles (which typically require 3 or more species) are absent, and removing each resident species from these stable communities is sufficient to construct the complete invasion graph. While starting with all species and allowing natural assembly could theoretically miss alternative stable states, our exploratory simulations confirmed that alternative states and priority effects are extremely rare in our system. Therefore, our approach preserves the key elements of \citepos{spaak2023building} framework: examining invasion growth rates across relevant subcommunities and ensuring an acyclic invasion graph. The primary advantage of our approach is computational tractability; with fewer subcommunities to analyze, we can measure growth rates with long simulations, across many posterior samples. 

\section{Why are some storage effects larger than others?}
\label{Decomposing Storage Effects}

\subsection{The components of a large storage effect}

To understand why certain sub-storage effects are larger than others, we must examine the fundamental components that contribute to storage effects. The storage effect tends to be large when 1) species have environmental niche differences, 2) environmental variation has substantial effects on per capita growth rates, 3) A good environment leads to high contemporaneous competition, and 4) there is subadditivity -- a negative interaction effect of environment and competition on per capita growth rates. Remarkably, these four features emerge in analytical expressions of the storage effect. For example, the storage effect (denoted $\Delta I$ in classic MCT) can be written as:
\begin{equation} \label{eq:components}
\Delta I \approx \overbrace{(1-\rho)}^{\text{Env. niche differences}} \times \underbrace{\sigma^2}_{\text{Env. var}} \times \overbrace{\theta}^{\text{Env. effect on Comp.}} \times \underbrace{(-\zeta)}_{\text{Env.-Comp. interaction}}
\end{equation}
This approximation (derived in \citealp{johnson2022storage} Appendix B) assumes two symmetric species --- species are equivalent except for the fact that they have partly uncorrelated responses to the environment (hence the lack of species-specific subscripts). While this mathematical expression from classic MCT does not apply directly to our analysis (a simulation-based analysis of a multi-species system with asymmetric species) it can nevertheless provide heuristic insight into the factors that determine storage effect strength.

To quantify these factors, we first define the \textit{standardized parameters} \citep{chesson1994multispecies}, which express environmental and competition in the common currency of per capita growth rates:

\begin{equation}
\mathscr{E}_j = r_j(E_j, \overline{C_j})
\end{equation}

\begin{equation}
\mathscr{C}_j = -r_j(\overline{E_j}, C_j)
\end{equation}

The components of the storage effect are then:

\begin{itemize}
\item Environmental niche differences ($1-\rho$): where $\rho = \text{Cor}(\mathscr{E}_j, \mathscr{E}_k)$ measures correlation between species' environmental responses
\item Environmental variance ($\sigma^2$): computed as $\text{Var}(\mathscr{E}_j)$.

\item Environmental effect on competition ($\theta$): measured as a partial derivative evaluated at equilibrium conditions:
\begin{equation}
\theta_j = \frac{\partial h_j(\myvect{\mathscr{E}}, \myvect{n})}{\partial \mathscr{E}_r} \Bigg|_{\substack{\myvect{\mathscr{E}}=\myvect{0} \\ \myvect{n}=\myvect{n^*}}},
\end{equation}
where is $h_j$ is a function that produces $\mathscr{C}_j$.

\item Environment-competition interaction ($\zeta$): quantified by the second partial derivative
\begin{equation}
\zeta_j = \frac{\partial^2 r(\mathscr{E}_j, C_j)}{\partial \mathscr{E}_j \partial \mathscr{C}_j} \Bigg|_{\substack{\mathscr{E}_j=0 \\ \mathscr{C}_j=0}}.
\end{equation}
\end{itemize}
Note that \citet{johnson2022storage} derive equation \ref{eq:components} using the non-standard parameters. However, under the small-noise assumptions of MCT, the two expressions can easily be converted to one another, e.g., via $\mathcal{E}_j =  \frac{\partial r(\mathscr{E}_j, C_j)}{\partial \mathscr{E}_j} \Bigg|_{\substack{\mathscr{E}_j=0 \\ \mathscr{C}_j=0}} \, (E_j-E_j^*). + \mathcal{O}(\epsilon^2)$, where $\epsilon$ is a small-noise parameter.

\subsection{Application to Sub-storage Effects}
\label{Application to Sub-storage Effects}

We seek to adapt the classic MCT decomposition to analyze our sub-storage effects. We reparameterize the per capita growth rate function to explicitly include survival, growth, and fecundity fluctuations: $r_j(S_j, G_j, F_j, C_j)$.
For each environmental component, we define standardized parameters by allowing one component to vary while holding others at their mean values. For example, the standardized fecundity parameter is:
\begin{equation}
\mathcal{F}_j = r_j(\overline{S_j}, \overline{G_j}, F_j, \overline{C_j})
\end{equation}
Similar expressions define $\mathcal{S}_j$ and $\mathcal{G}_j$ for survival and growth respectively. The standardized competition parameter is defined with a negative sign so that higher competition corresponds to smaller growth rates:
\begin{equation}
\mathcal{C}_j = -r_j(\overline{S_j}, \overline{G_j}, \overline{F_j}, C_j)
\end{equation}

To quantify the components of each sub-storage effect, we analyzed communities where at least two species could coexist. For each such community, we simulated 1000 years of dynamics after a short burn-in period, calculated standardized parameters for each species, and computed the following quantities:
\begin{itemize}
\item Environmental (between-species) correlations, $(1-\rho')$: Calculated all pairwise correlations (e.g., $\text{Cor}(\mathcal{S}_i, \mathcal{S}_j)$) for each environmental parameter

\item Environmental variance, $\sigma^{\prime 2}$: Computed $\text{Var}(\mathcal{S}_j)$, $\text{Var}(\mathcal{G}_j)$, and $\text{Var}(\mathcal{F}_j)$

\item Environmental effect on competition, $\theta'$ : Performed linear regressions with $\mathcal{C}_j$ as response and each standardized environmental parameter as predictor

\item Environment-competition interaction, $\zeta'$ : Fit multiple regressions with interactions between $\mathcal{C}_j$ and each environmental parameter. The coefficients of the interaction terms were identified as the environment-competition interaction effects.

\item The approximate storage effect: for each species pair and each standardized environmental parameter, we computed the analogue of \eqref{eq:components}, e.g., $(1-\rho') \times \sigma^{\prime2}  \times \theta' \times \left(- \zeta' \right)$. 
\end{itemize}

The prime notation (') indicates that these computed terms are approximations of their theoretical counterparts defined in \eqref{eq:components}. Exploratory simulations confirmed approximately linear relationships between standardized variables, allowing us to use linear regression as a practical method for estimating partial derivatives.

% latex table generated in R 4.4.1 by xtable 1.8-4 package
% Wed Jan 15 18:01:02 2025
\begin{table}
\begin{tabular}{l|ccc}
  \toprule
Effect Type & S & G & F \\ 
  \midrule
  % Env. var., $\sigma^{\prime 2}$ & $8.28 \cdot 10^{-3}$ & $5.64 \cdot 10^{-2}$ & $7.40 \cdot 10^{-2}$ \\ 
  % Env. effect on Comp., $\theta'$ & -2.68 & 0.30 & 1.15 \\ 
  % Env.-Comp. interaction effect, $\zeta'$ & 2.45 & -0.41 & -2.91 \\ 
  % Approx. storage effect & $1.97 \cdot 10^{-2}$ & $3.83 \cdot 10^{-2}$ & $1.03 \cdot 10^{-1}$ \\ 
  Env. correlations, $\rho'$ & 0.19 & 0.08 & -0.07 \\ 
  Env. var., $\sigma^{\prime 2}$ & $7.40 \cdot 10^{-3}$ & $5.20 \cdot 10^{-2}$ & $6.60 \cdot 10^{-2}$ \\ 
  Env. effect on Comp., $\theta'$ & -3.34 & 0.29 & 1.21 \\ 
  Env.-Comp. interaction effect, $\zeta'$ & 2.41 & -0.27 & -2.78 \\ 
  Approx. storage effect & $1.94 \cdot 10^{-2}$ & $8.13 \cdot 10^{-3}$ & $9.20 \cdot 10^{-2}$ \\ 
   \bottomrule
\end{tabular}
\caption{Storage effect component decomposition, for survival (S), growth (G), and fecundity (F). For each community with $\geq2$ coexisting species, components were first averaged across species pairs, then the median was taken across all posterior parameter sets.} 
\label{tab:classic_decomp}
\end{table}

This decomposition of sub-storage effects into their component parts (Table \ref{tab:classic_decomp}) reveals several insights. Fecundity shows the strongest storage effect because fecundity fluctuations have a large effect on per capita growth rates (i.e.,  $\sigma^{\prime 2}$ is large in the ``F'' column). Additionally, the interaction effect between fecundity and competition is large and negative, i.e., coexistence-promoting. 

\section{Fecundity fluctuations in the literature}
\label{Fecundity fluctuations in the literature}

In the main text, we argued that the storage effect is weak not just in our study system, but in reef-building corals more generally. Here we examine one such justification for this claim: fecundity fluctuations --- the primary driver of the storage effect in our model --- are larger in our model than in most coral systems. Since our model generates a weak storage effect with atypically large fecundity fluctuations, we expect the storage effect to be even weaker in nature, on average, for Indo-Pacific coral reefs. While a comprehensive analysis of all storage effect elements (i.e., the rows of Table \ref{tab:classic_decomp}) across coral communities would be ideal, such data are not currently available. The literature does, however, contain some data on fecundity fluctuations.

We identified six studies that contained multi-year coral fecundity data \citep{hughes2000supply, borger2010effects, st2016fecundity, tan2016spatial, pratchett2019spatial, alessi2024spawning}. Most of the data were extracted from published tables and figures, but \citet{hughes2000supply} provided their data upon request.  We specifically focused on two metrics: the probability that a colony is reproductive (PRR), and/or the number of oocytes per polyp, conditioned on the colony being reproductive (OPP). These two quantities are predicted by our model (see \eqref{eq:reproductive1} \& \eqref{eq:log_eggs1}), allowing a fair model-to-literature comparison. We believe these six studies represent the majority of publicly available multi-year fecundity datasets, but an exhaustive literature review was not possible due to the large size of the literature and differences in terminology across studies.

To generate comparable predictions from our model, we implemented a systematic simulation protocol. From the joint posterior distribution, we randomly selected 100 parameter sets. For each parameter set, we chose one species at random and simulated its population dynamics in isolation for 150 years. After discarding the first 50 years to eliminate transient dynamics, we calculated PRR and OPP for colonies exceeding a planar area threshold. Several planar area thresholds were used: 10 cm\textsuperscript{2} (the rule-of-thumb minimum size for species identification), 63 cm\textsuperscript{2} (the minimum size in \citealp{st2016fecundity}), 201 cm\textsuperscript{2} (the minimum size in \citealp{tan2016spatial}), and 707 cm\textsuperscript{2} (the minimum size in \citealp{hughes2000supply}). When studies reported only diameter thresholds, we calculated the corresponding planar areas assuming circular colony shapes. Some studies in our analysis lacked clearly defined size thresholds for sampling.

We analyzed both simulated and empirical data using three summary statistics for reproductive metrics (PRR \& OPP): the mean, standard deviation (SD), and coefficient of variation (CV). For our simulations, we calculated these statistics for each run and report the median (across simulations) in Tables \ref{tab:PRR} and \ref{tab:OPP}. The SD directly quantifies the magnitude of reproductive fluctuations, while the CV normalizes these fluctuations by the mean, putatively correcting for differences in fitness across species.

All six studies provide two years of data. To estimate the SD and CV at all, we therefore aggregated data across all available dimensions: years, species, and locations. While this aggregation inflates both SD and CV by combining multiple sources of variation, our IPM-generated fecundity data still showed higher variability in several comparisons. Specifically, simulated fecundity exhibited higher coefficients of variation for PRR compared to half of the empirical studies (2 out of 4), regardless of the minimum colony area used (Table \ref{tab:PRR}).

\begin{figure}[H]
\centering
\makebox[\textwidth]{\includegraphics[scale=0.9]{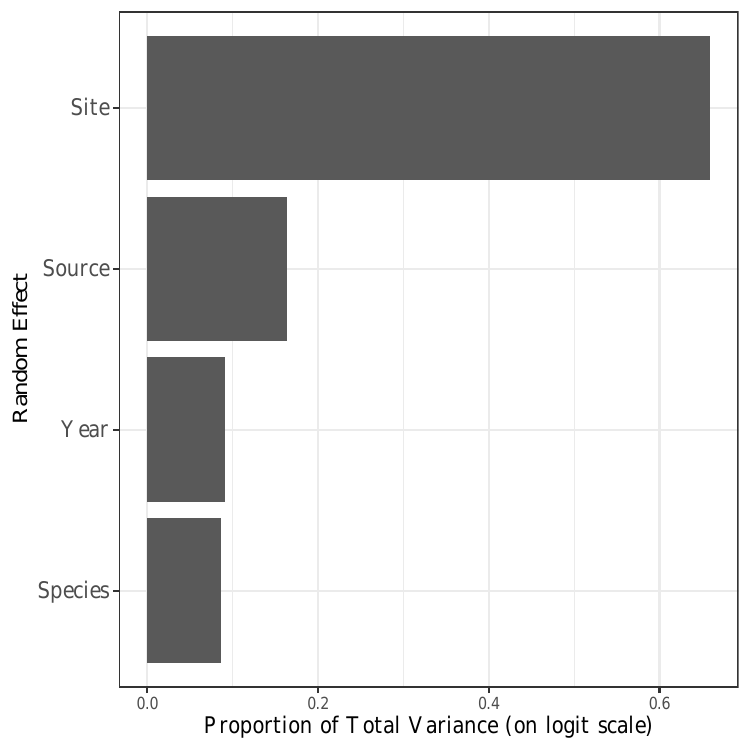}}
\caption{Decomposition of variance in the probability of being reproductive (PRR), as estimated by the beta-link random effects model. The analyzed variable is PRR after boundary compression and logit transformation: $\text{logit}\left( (\text{PRR} * (n-1) + 0.5) / n \right)$, where n is the total number of PRR estimates across sites, source, year, and species.}
\label{fig:var_partition}
\end{figure}

We attempted to isolate purely temporal variation in fecundity metrics by statistically adjusting the SD and CV values reported in the literature. For the probability of being reproductive (PRR), we developed a generalized linear mixed model using a beta link function, with random effects for species, site, study, and year. We implemented this model using the \textit{glmmTMB} package \citep{glmmTMB} and evaluated its adequacy \textit{DHARMa} package \citep{dharma2024} diagnostics. The model reveals that temporal variation accounts for only 9\% of PRR variation, while spatial variation explains 66\% (Fig. \ref{fig:var_partition}).

To adjust the literature-derived statistics, we simulated PRR values using two versions of the generalized linear mixed model: one incorporating all sources of variation and another including only temporal variation. By calculating the ratio of simulated CVs between these models, we derived a multiplicative correction factor. We applied this factor to the CVs from the literature to produce the adjusted SD and CV values shown in the right-most columns of Table \ref{tab:OPP}. Measured with these adjusted CVs, fecundity fluctuations are almost always greater in the IPM-based simulations than in the literature; the exception is that \citet{hughes2000supply}, and Simulations with a 707 cm\textsuperscript{2} threshold, produce CVs of 0.116 and 0.100 respectively.

We were unable to perform similar adjustments for OPP due to statistical challenges; the data contained many outliers, and we could not identify a suitable transformation or link function that produced satisfactory diagnostics. Consequently, Table \ref{tab:OPP} presents only unadjusted SD and CV values for OPP.

% latex table generated in R 4.4.1 by xtable 1.8-4 package
% Mon Jan 20 18:27:50 2025
\begin{table}[H]
\caption{Simulation vs. literature comparison of variation in the probability of being reproductive (PRR). The adjusted SD and CV aim to isolate temporal variation; see Appendix text for computational details. Studies that did not specify a minimum colony size for sampling are marked as N/A in the 'Min colony area' column.} \label{tab:PRR}
\centering
\begin{tabular}{lrrrrrr}
  \hline
Source & \makecell{Min. colony \\area cm\textsuperscript{2}} & Mean & SD & CV & \makecell{SD\\(adjusted)} & \makecell{CV\\(adjusted)} \\ 
  \hline
Simulations & 10 & 0.341 & 0.168 & 0.494 & 0.168 & 0.494 \\ 
  Simulations & 63 & 0.610 & 0.157 & 0.313 & 0.157 & 0.313 \\ 
  Simulations & 201 & 0.814 & 0.110 & 0.147 & 0.110 & 0.147 \\ 
  Simulations & 707 & 0.938 & 0.094 & 0.102 & 0.094 & 0.102 \\ 
  \citet{alessi2024spawning} & N/A & 0.961 & 0.049 & 0.051 & 0.014 & 0.014 \\ 
  \citet{borger2010effects} & N/A & 0.687 & 0.165 & 0.240 & 0.048 & 0.065 \\ 
  \citet{hughes2000supply} & 707 & 0.670 & 0.286 & 0.427 & 0.082 & 0.116 \\ 
  \citet{tan2016spatial} & 201 & 0.940 & 0.053 & 0.056 & 0.015 & 0.015 \\ 
   \hline
\end{tabular}
\end{table}

% latex table generated in R 4.4.1 by xtable 1.8-4 package
% Mon Jan 20 19:24:29 2025
\begin{table}[H]
\caption{Simulation vs. literature comparison of the variation in oocytes per polyp (OPP).} \label{tab:OPP} \centering
\begin{tabular}{lrrrr}
  \hline
Source & \makecell{Min. colony \\area cm\textsuperscript{2}} & Mean & SD & CV \\ 
  \hline
Simulations & 10 & 1.383 & 0.845 & 0.544 \\ 
  Simulations & 63 & 2.481 & 1.006 & 0.399 \\ 
  Simulations & 201 & 3.634 & 0.950 & 0.224 \\ 
  Simulations & 707 & 5.242 & 1.045 & 0.211 \\ 
  \citet{alessi2024spawning} & N/A & 8.803 & 1.584 & 0.180 \\ 
  \citet{borger2010effects} & N/A & 66.733 & 21.204 & 0.318 \\ 
  \citet{pratchett2019spatial} & N/A & 5.839 & 1.569 & 0.269 \\ 
  \citet{st2016fecundity} & 63 & 230.468 & 59.136 & 0.257 \\ 
  \citet{tan2016spatial} & 201 & 6.153 & 1.291 & 0.210 \\ 
   \hline
\end{tabular}
\end{table}

\section{Supplementary figures and tables} \label{supplementary figures}

\begin{figure}[H]
\centering
\includegraphics[scale = 1]{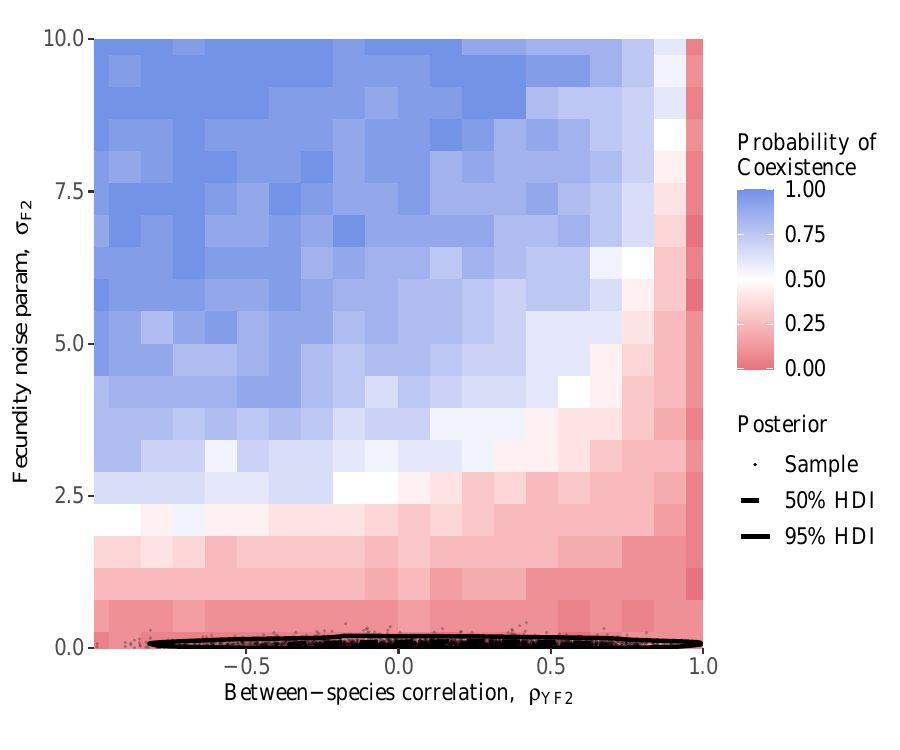}
\caption{Coexistence via the storage effect occurs when species have strong environmental niche differences (low $\rho_{F2}$) combined with high environmental variation (high $\sigma_{j,YF2}$), but our fitted model indicates that species fall far outside this regime. The heatmap color shows the posterior probability of coexistence between two empirically-codominant species, \textit{A. digitifera} and \textit{A. hyacinthus}, as a function of two parameters: the between-species correlation ($\rho_{F2}$) and the scale of environmentally-driven variation ($\sigma_{j,YF2}$) in coral fecundity. Specifically, we modulate the egg densities conditional on colonies being gravid, which differentiates this figure from figure \ref{fig:high_var_tiles}, in which we modulated the probability of being gravid. Posterior probabilities of coexistence were calculated from 30 model simulations with fixed $\rho_{F2}$ and $\sigma_{j,YF2}$. Marginal distributions of these parameters are shown by posterior samples and high-density intervals (HDI).}
\label{fig:high_var_tiles_F2}
\end{figure}

\begin{figure}[H]
\centering
\includegraphics[scale = 1]{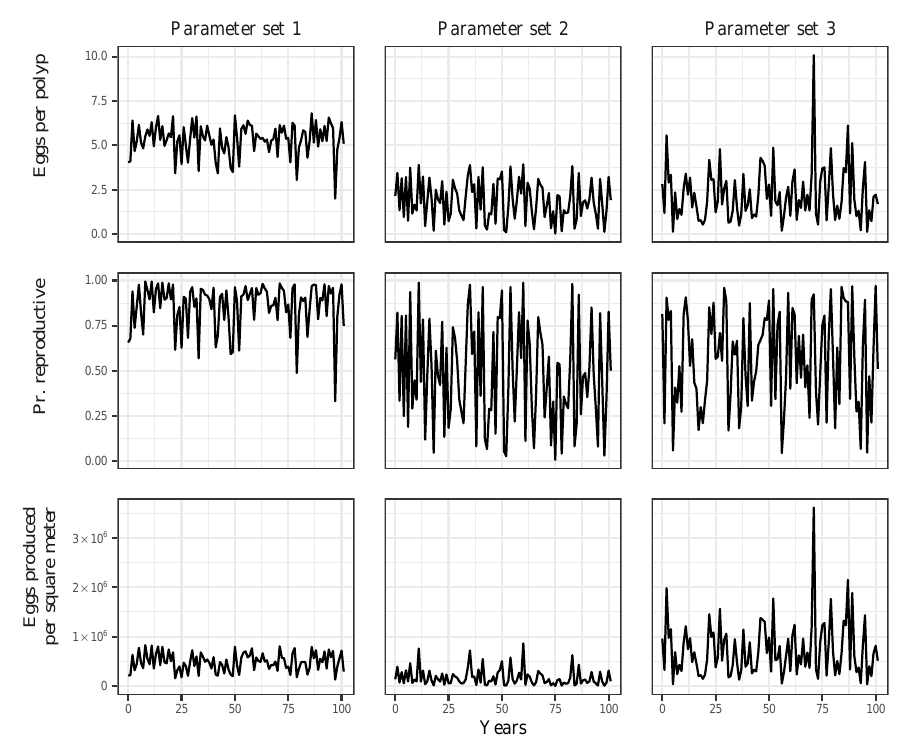}
\caption{Simulated time series of fecundity metrics (rows) for three randomly selected parameter sets (columns). Simulations used the ``baseline'' rules (i.e., not any of the alternative model structures in Appendix \ref{Robustness analysis}), and contained a single, randomly selected species.}
\label{fig:fecundity_var_TS}
\end{figure}

\begin{figure}[H]
\centering
\includegraphics[scale = 1]{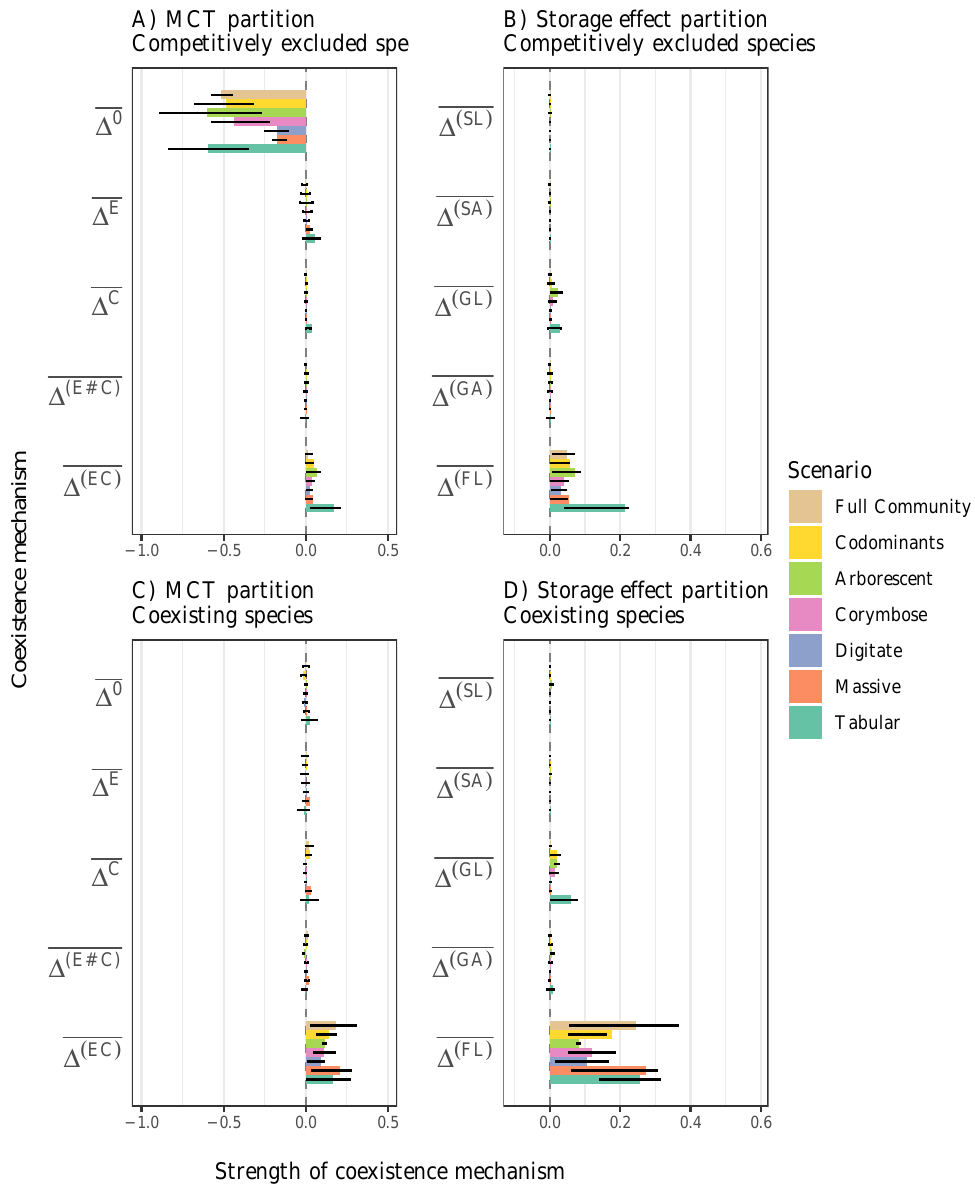}
\caption{Coexistence mechanism strength is relatively consistent, regardless of the community type. Panels show coexistence mechanism strength across competitively excluded species (top row), coexisting species (bottom row), the coarse-grained MCT partition (left column), and the fine-grained MCT partition showing sub-storage effects (right column). Bars and error bars respectively show the posterior means and interquartile ranges. Data were generated with the baseline IPM simulations.}
\label{fig:MCT_partition_scenarios}
\end{figure}

\begingroup
\small 
\begin{longtable}[H]{p{2.2cm} p{7.4cm} p{3.3cm} p{2.2cm}} 
\caption{Comprehensive table of model symbols, including the wave-disturbance, colony growth, fecundity, recruitment sub-models, the integral projection model, and modern coexistence theory (MCT) partitions.  ``Units'' is ``dimensionless'' where no strict physical unit applies.  The ``Type'' column differentiates fixed versus estimated parameters, empirical observations, and derived quantities. Posterior draws can be found the in the supplementary files: {\fontfamily{qcr}\selectfont $\sim$ /coral\_sims\_HPCC/demographic\_pars.rds}} \label{tab:symbols} \\
\toprule
\textbf{Symbol} & \textbf{Meaning} & \textbf{Units} & \textbf{Type} \\
\midrule
\endfirsthead

\multicolumn{4}{c}{{\tablename} \thetable{} -- continued from previous page} \\
\toprule
\textbf{Symbol} & \textbf{Meaning} & \textbf{Units} & \textbf{Type} \\
\midrule
\endhead

\bottomrule
\endlastfoot

\multicolumn{4}{l}{\textbf{Wave-disturbance sub-model (Sec.\ \ref{Wave action survival})}} \\
\midrule
\(v_t\)           
  & Annual maximum wind velocity (drawn from Gamma)               
  & m\,s\(^{-1}\)   
  & Derived \\
\(\alpha_W\)      
  & Shape parameter for Gamma distribution of \(v_t\)             
  & dimensionless    
  & Fixed par. \\
\(\beta_W\)       
  & Scale parameter for Gamma distribution of \(v_t\)             
  & m\,s\(^{-1}\)    
  & Fixed par. \\
\(u_t\)           
  & Water velocity, computed via \(u = a(1-e^{-b v})\)             
  & m\,s\(^{-1}\)     
  & Derived \\
\(a\)             
  & Water velocity parameter              
  & m\,s\(^{-1}\)     
  & Fixed par. \\
\(b\)             
  & Water velocity parameter                     
  & s\,m\(^{-1}\)     
  & Fixed par. \\
\(\rho_w\)        
  & Seawater density                                               
  & kg\,m\(^{-3}\)    
  & Fixed par. \\
\(\sigma_s\)      
  & Substrate tensile strength                                     
  & MN\,m\(^{-2}\)     
  & Fixed par. \\
\(\text{DMT}_t\) 
  & Dislodgment mechanical threshold                               
  & dimensionless     
  & Derived \\

\midrule
\multicolumn{4}{l}{\textbf{Colony shape factor (CSF)}} \\
\midrule
\(\text{CSF}\)    
  & Colony shape factor \(\frac{16}{d_{\parallel} d_{\perp}}\!\int_0^{h} w(y)\,dy\) 
  & dimensionless     
  & Derived \\
\(d_{\parallel},\, d_{\perp}\)   
  & Perpendicular basal attachment widths of the colony            
  & m                  
  & Observed \\
\(h\)             
  & Maximum colony height above substrate                          
  & m                  
  & Observed \\
\(w(y)\)          
  & Colony width at height \(y\)                                   
  & m                  
  & Observed \\

\midrule
\multicolumn{4}{l}{\textbf{CSF regression model}} \\
\midrule
\(x_i\)        
  & \(\log\bigl(\text{planar area [cm}^2]\bigr)\) for colony \(i\) 
  & dimensionless   
  & Observed \\
\(\beta_{m[i],0C},\, \beta_{m[i],1C}\) 
  & Morphology-specific intercept/slope for \(\log(\text{CSF})\)   
  & dimensionless    
  & Estim. par. \\
\(\sigma_{m,C}\)  
  & Morphology-specific residual SD in \(\log(\text{CSF})\)        
  & dimensionless    
  & Estim. par. \\
\(\widehat{\text{CSF}}_j(x)\)    
  & Model-predicted CSF for species \(j\) at size \(x\)            
  & dimensionless    
  & Derived \\
\(D_{j,t}(x)\)    
  & Disturbance-driven mortality indicator (0 or 1)                
  & dimensionless    
  & Derived \\

\midrule
\multicolumn{4}{l}{\textbf{Background mortality (Sec.\ \ref{Background mortality})}} \\
\midrule
\(\text{surv}_{i}\)  
  & Binary survival indicator (1 if alive, 0 if dead)   
  & dimensionless    
  & Observed \\

\(\beta_{j[i],0S},\,\beta_{j[i],1S}\)  
  & Species-specific logistic intercept/slope for \(\log(\text{area})\)  
  & dimensionless    
  & Estim. par. \\
\(M_j(x)\)        
  & Fraction of colonies (size \(x\)) that die (background only)    
  & dimensionless    
  & Derived \\
\(S_{j,t}(x)\)    
  & Annual survival probability \((1 - D_{j,t}(x))(1 - M_j(x))\)    
  & dimensionless    
  & Derived \\

\midrule
\multicolumn{4}{l}{\textbf{Colony growth (Sec.\ \ref{Colony growth})}} \\
\midrule
\(x_i,\, x'_i\)  
  & \(\log(\text{colony area [cm}^2])\) in current vs.\ next year   
  & dimensionless    
  & Observed \\
\(G_i = \frac{\exp(x'_i)}{\exp(x_i)}\)  
  & Colony growth ratio (raw-area quotient)                         
  & dimensionless    
  & Derived \\
\(g_i\)          
  & Box-Cox transformed growth ratio                               
  & dimensionless    
  & Derived \\
\(\lambda\)       
  & Box-Cox exponent (fitted via MLE)                              
  & dimensionless    
  & Estim. par. \\
\(\nu_j\)         
  & Species-specific degrees of freedom (Student-\(t\))             
  & dimensionless    
  & Estim. par. \\
\(\sigma_j\)      
  & Species-specific residual SD (Student-\(t\))                    
  & dimensionless    
  & Estim. par. \\
\(\beta_{j,0G},\, \beta_{j,1G}\) 
  & Species-specific intercept/slope for colony growth model        
  & dimensionless    
  & Estim. par. \\
\(\eta_{j,t,G}\)  
  & Species-specific random year effect (growth)                    
  & dimensionless    
  & Estim. par. \\
\(\sigma_{j,YG}\) 
  & SD of year effects (growth)                                    
  & dimensionless    
  & Estim. par. \\
\(\rho_{j,k,G}\)  
  & Correlation among species \(j\) and \(k\) in growth year effects
  & dimensionless    
  & Derived \\
\(\mathbf{R}_G\)  
  & Correlation matrix for growth year effects across species       
  & dimensionless    
  & Derived \\
\(\mathbf{\Sigma}_G\)  
  & Covariance matrix for species’ growth year effects (MVN)        
  & dimensionless    
  & Derived \\
\(\boldsymbol{\eta}_{t,G}\) 
  & Vector of species-specific growth year effects in year \(t\)    
  & dimensionless    
  & Derived \\
\(G_{j,t}(y \mid x)\)  
  & Growth-transition kernel from size \(x\) to \(y\)               
  & dimensionless    
  & Derived \\

\midrule
\multicolumn{4}{l}{\textbf{Fecundity sub-model (Sec.\ \ref{Fecundity})}} \\
\midrule
\(\text{reproductive}_{i}\)   
  & Binary indicator (1 if colony \(i\) produced eggs)              
  & dimensionless    
  & Observed \\
\(\gamma_{j,0F},\, \gamma_{j,1F}\) 
  & Species-specific intercept/slope for reproductive probability   
  & dimensionless    
  & Estim. par. \\
\(\eta_{j,t,F1}\)  
  & Species-specific year effect (reproductive probability)  
  & dimensionless    
  & Estim. par. \\
\(\sigma_{j,YF1}\) 
  & SD of year effects  (reproductive probability)              
  & dimensionless    
  & Estim. par. \\
\(\rho_{j,k,F1}\)  
  & Correlation of year effects (reproductive probability) between species \(j\) and \(k\)  
  & dimensionless    
  & Derived \\
\(\mathbf{R}_{F1}\)  
  & Correlation matrix for species’ year effects (reproductive probability) for MVN
  & dimensionless    
  & Derived \\
\(\mathbf{\Sigma}_{F1}\)  
  & Covariance matrix for species’ year effects (reproductive probability) for MVN
  & dimensionless    
  & Derived \\
\(\boldsymbol{\eta}_{t,F1}\) 
  & Vector of species-specific fecundity year effects (reproductive probability)  
  & dimensionless    
  & Derived \\
\(\text{eggs}_{i}\) 
  & Observed mean \# of eggs per polyp for colony \(i\)             
  & \#\,eggs\,polyp\(^{-1}\) 
  & Observed \\
\(\beta_{j,0F},\, \beta_{j,1F}\) 
  & Intercept/slope for \(\log(\text{eggs per polyp})\) (skew-normal location) 
  & dimensionless    
  & Estim. par. \\
\(\eta_{j,t,F2}\)  
  & Species-specific year effect (log eggs per polyp)            
  & dimensionless    
  & Estim. par. \\
\(\sigma_{j,YF2}\) 
  & SD of year effects (log eggs per polyp)                             
  & dimensionless    
  & Estim. par. \\
\(\rho_{j,k,F2}\)  
  & Correlation of year effects (log eggs per polyp) between species \(j\) and \(k\)  
  & dimensionless    
  & Derived \\
  \(\mathbf{R}_{F2}\)  
  & Correlation matrix for species’ year effects (log eggs per polyp) for MVN
  & dimensionless    
  & Derived \\
\(\mathbf{\Sigma}_{F2}\)  
  & Covariance matrix for species’ year effects (log eggs per polyp) for MVN
  & dimensionless    
  & Derived \\
\(\boldsymbol{\eta}_{t,F2}\) 
  & Vector of species-specific year effects (log eggs per polyp) 
  & dimensionless    
  & Derived \\
\(\omega_j\)      
  & Scale parameter (skew-normal)                                   
  & dimensionless    
  & Estim. par. \\
\(\alpha_j\)      
  & Shape parameter (skew-normal)                                   
  & dimensionless    
  & Estim. par. \\
\(\theta_j\)      
  & Mean polyp density (partially pooled across species)            
  & \#\,polyps\,cm\(^{-2}\) 
  & Estim. par. \\
\(\mu_\theta,\;\sigma_\theta\) 
  & Hyperparameters for polyp density                              
  & \#\,polyps\,cm\(^{-2}\) 
  & Estim. par. \\
\(p_{j,t}(x)\)     
  & Probability colony is reproductive, given size \(x\)            
  & dimensionless    
  & Derived \\
\(\xi_{j,t}(x)\)   
  & Location parameter in \(\log(\text{eggs})\) skew-normal         
  & dimensionless    
  & Derived \\
\(F_{j,t}(x)\)     
  & Colony-level fecundity (total \#\,eggs)                         
  & \#\,eggs        
  & Derived \\

\midrule
\multicolumn{4}{l}{\textbf{Recruitment sub-model (Sec.\ \ref{Recruitment})}} \\
\midrule
\(L_{j,t}\) 
  & Total egg production of species \(j\) per m\(^2\)              
  & \#\,eggs\,m\(^{-2}\) 
  & Derived \\
\(N'_{j,t} \)  
  & Proportion of area occupied by species \(j\) 
    (after survival/growth)                                        
  & dimensionless    
  & Derived \\
\(R_{j,t+1}\)   
  & Density of one-year recruits (species \(j\))                    
  & \#\,recruits\,m\(^{-2}\) 
  & Derived \\
\(\beta_{j,R}\)    
  & Maximum recruit density parameter (tuned via simulation)        
  & dimensionless    
  & Derived \\
\(\mu_D, \;\sigma_D\)  
  & Mean, SD of recruit diameter (normal dist.)                     
  & cm                
  & Fixed par. \\
\(\phi_j(x)\)     
  & Size distribution (PDF) of new recruits over log-area bins      
  & dimensionless    
  & Derived \\

\midrule
\multicolumn{4}{l}{\textbf{Integral Projection Model (Sec.\ \ref{Complete model formulation})}} \\
\midrule
\(n_{j,t}(x)\)     
  & Density of colonies of species \(j\) at size \(x\), time \(t\)  
  & \#\,colonies\,m\(^{-2}\) 
  & Derived \\
\(n_{j,t+1}(y)\)  
  & Next-year density at size \(y\), after survival/growth          
  & \#\,colonies\,m\(^{-2}\) 
  & Derived \\
\(S_{j,t}(x)\)     
  & Annual survival probability                                      
  & dimensionless    
  & Derived \\
\(D_{j,t}(x)\)     
  & Disturbance-induced mortality (1 or 0)                          
  & dimensionless    
  & Derived \\
\(M_j(x)\)         
  & Background mortality fraction                                   
  & dimensionless    
  & Derived \\

\midrule
\multicolumn{4}{l}{\textbf{Population growth rate (Sec.\ \ref{Complete model formulation})}} \\
\midrule
\(\lambda_{j,t}\)  
  & Finite rate of increase for species \(j\) at time \(t\)         
  & dimensionless    
  & Derived \\
\(r_{j,t}\)   
  & Per capita growth rate $ = \log(\lambda_{j,t})$                                          
  & dimensionless    
  & Derived \\
\end{longtable}
\endgroup

\end{appendices}

\bibliographystyle{apalike}
\bibliography{refs.bib}

\end{document}